\documentclass{IEEEtran}
\IEEEoverridecommandlockouts
\usepackage{cite}
\usepackage{float}
\usepackage{stfloats}
\usepackage{fancyhdr}
\usepackage{color}
\usepackage{amsmath}
\usepackage{graphicx} 
\graphicspath{{./figures/}}
\usepackage{algorithm}
\usepackage{algorithmic}
\usepackage{epsfig}
\usepackage{subfigure}
\usepackage{amssymb}
\usepackage{empheq}  
\usepackage{extarrows}
\usepackage{longtable}
\usepackage{booktabs}
\usepackage{diagbox}

\usepackage{url}
\usepackage[hidelinks]{hyperref}

\usepackage{amsthm}

\usepackage{etoolbox}
\makeatletter
\def\@IEEEBIOskipN{1.5\baselineskip}
\expandafter\patchcmd\csname \string\IEEEbiography\endcsname
{\vskip \@IEEEBIOskipN plus 1fil minus 0\baselineskip}
{\vskip \@IEEEBIOskipN}
{\typeout{*** IEEE biography spacing patch SUCCESS ***}}
{\typeout{*** IEEE biography spacing patch FAILED ***}}
\makeatother

\usepackage{txfonts}
\usepackage{bm}
\begin{document}
	\bstctlcite{IEEEexample:BSTcontrol}
	
	\title{Environment-Aware Channel Inference via Cross-Modal Flow: From Multimodal \\Sensing to Wireless Channels}
	
	\author{Guangming~Liang,~\IEEEmembership{Graduate Student Member,~IEEE}, Mingjie~Yang,~\IEEEmembership{Graduate Student Member,~IEEE}, Dongzhu~Liu,~\IEEEmembership{Member,~IEEE}, Paul Henderson, and Lajos Hanzo,~\IEEEmembership{Life Fellow,~IEEE}
	\thanks{The financial support of the following Engineering and Physical Sciences Research Council (EPSRC) projects is gratefully acknowledged: Platform for Driving Ultimate Connectivity (TITAN) (EP/X04047X/1; EP/Y037243/1); Robust and Reliable Quantum Computing (RoaRQ, EP/W032635/1); PerCom  EP/X01228X/1; EP/Y026721/1, India-UK Intelligent Spectrum Innovation ICON UKRI-1859. This work was presented in part at IEEE International Conference on Communications (ICC), Glasgow, U.K., May 2026 \cite{ICC_Early_Version}.}
	\thanks{Guangming Liang, Mingjie Yang, Dongzhu Liu and Paul Henderson are with the School of Computing Science, University of Glasgow, Glasgow G12 8QQ, U.K., e-mails: \{3032221l,2921021y\}@student.gla.ac.uk, \{Dongzhu.Liu, Paul.Henderson\}@glasgow.ac.uk. \textit{(Corresponding author: Dongzhu Liu.)}}
    \thanks{Lajos Hanzo is with the School of Electronics and Computer Science, University of Southampton, Southampton SO17 1BJ, U.K., e-mail: lh@ecs.soton.ac.uk.}
    }

	\maketitle
	
	\begin{abstract}
   		Accurate channel state information (CSI) underpins reliable and efficient wireless communication. However, acquiring CSI via pilot estimation incurs substantial overhead, especially in massive multiple-input multiple-output (MIMO) systems operating in high-Doppler environments. By leveraging the growing availability of environmental sensing data, this treatise investigates pilot-free channel inference that estimates complete CSI directly from multimodal observations, including camera images, LiDAR point clouds, and GPS coordinates. In contrast to prior studies that rely on predefined channel models, we develop a data-driven framework that formulates the sensing-to-channel mapping as a cross-modal flow matching problem. The framework fuses multimodal features into a latent distribution within the channel domain, and learns a velocity field that continuously transforms the latent distribution toward the channel distribution. To make this formulation tractable and efficient, we reformulate the problem as an equivalent conditional flow matching objective and incorporate a modality alignment loss, while adopting low-latency inference mechanisms to enable real-time CSI estimation. In experiments, we build a procedural data generator based on Sionna and Blender to support realistic modeling of sensing scenes and wireless propagation. System-level evaluations demonstrate significant improvements over pilot- and sensing-based benchmarks in both channel estimation accuracy and spectral efficiency for the downstream beamforming task. The source code is available at \url{https://github.com/gm-leung/environment-aware-channel-inference}.
    \end{abstract}
	
	
	\section{Introduction}
    As wireless networks evolve toward next generation (NG) systems, they are expected to support emerging applications such as immersive extended reality, digital twins, and holographic communications \cite{10054381}. These applications require highly reliable yet low-latency communication links, which rely on accurate knowledge of the channel state information (CSI) to fully exploit the potential of massive multiple-input multiple-output (MIMO) technology \cite{10379539}. However, acquiring high-resolution CSI for large-scale antenna arrays remains a major bottleneck, as traditional channel estimation methods introduce pilot overhead that scales with the number of antennas and user terminals, thereby occupying spectral resources that could otherwise be used for data transmission. In simple tangible terms, the pilot overhead must be doubled every time the Doppler frequency is doubled. Consequently, significant research efforts have been devoted to reducing pilot overhead without sacrificing channel estimation accuracy \cite{7094443,6998861,10836149,10315065,10845822,10829589,9957135,10930691,10615282}. Despite these advances, achieving accurate and low- or even zero-overhead CSI acquisition remains a key challenge in unlocking the full potential of massive MIMO in NG networks.

    Environment-aware communications \cite{10430216} have recently emerged as a promising paradigm to reduce or even eliminate reliance on pilot signals for CSI acquisition. By sensing environmental features, such as obstacle locations, reflector and scatterer characteristics, terrain and building layouts, and dynamic objects, communication systems can either directly infer channel states or predict propagation models characterized by path loss, shadowing, multipath, and LoS/NLoS conditions. This environmental awareness enables optimized transmission strategies, including beamforming, user scheduling, and resource allocation, without requiring explicit pilot-based CSI estimation. Existing studies on environment-aware communications generally follow a pair of representative strategies. The first one leverages multimodal sensing data, including GPS coordinates, camera RGB images, LiDAR point clouds, and radar signals, to perform channel-related downstream tasks such as beamforming design \cite{10539181} and link-blockage detection \cite{10008524}. Although this task-oriented approach bypasses explicit CSI estimation, the learned mappings are typically task-specific and lack generalization across system configurations. The second strategy relies solely on GPS-based localization to infer CSI under the assumptions of quasi-static environments \cite{9473871,10.1145/3570361.3592527}, which is inherently inapplicable to dynamic propagation scenarios.  

    To address the limitations discussed above, we propose a flow-matching-based generative framework that learns a continuous transformation from multimodal environmental sensing data to high-dimensional wireless channel states. This represents the first environment-aware technique capable of achieving real-time full CSI recovery from dynamic multimodal sensing data without relying on pilot signaling. By leveraging computational intelligence to replace communication overhead, the proposed framework enhances spectral efficiency, supports physical-layer operations facilitated by full CSI, and provides low-latency inference suitable for practical deployment in dynamic NG networks.

	\subsection{Related Works} 
    \subsubsection{Channel Estimation with Reduced Pilot Overhead} To mitigate the pilot overhead in massive MIMO systems, model-based \cite{7094443,6998861} and data-driven \cite{10836149,10315065,10845822,10829589,9957135,10930691} methods have been proposed to reduce pilot usage, while preserving channel estimation accuracy. Model-based techniques, such as compressed sensing, exploit angular-domain sparsity to estimate high-dimensional CSI from limited pilot measurements \cite{7094443,6998861}. However, the sparsity assumption may not strictly hold in dynamic or rich-scattering environments, which limits their generalization capability. Data-driven methods aim to recover full CSI from partial estimates by capturing correlations across spatial, frequency, and temporal domains \cite{10836149,10315065} or by integrating auxiliary information such as historical CSI \cite{10845822} and multi-view images \cite{10829589}. Despite their effectiveness, these approaches often require large datasets and are typically tailored to specific pilot configurations, which limits their generalization to new scenarios. Recently, generative probabilistic models based on score matching \cite{9957135} and diffusion processes \cite{10930691,10615282} have been explored for learning the underlying channel distribution and estimate CSI through posterior sampling under arbitrary pilot settings. While these methods reduce pilot dependence, pilot-free CSI acquisition remains a radical open challenge, motivating the generative framework proposed in this paper. 

	\begin{table*}[t]  
		\caption{Contrasting our contributions to the literature of channel inference methods}
		\footnotesize
		\begin{center} 
			\begin{tabular}{|l|c|c|c|c|c|c|c|c|c|c|c|}
				\hline
				&\cite{7094443,6998861}  &\cite{10836149,10315065,10845822,10829589,9957135,10930691} & \cite{9473871} & \cite{10.1145/3570361.3592527} & \cite{wang2024multi,zhang2025vision,10720899}  & Our work \\ \hline
				Instantaneous CSI acquisition   &\checkmark  & \checkmark  &\checkmark&\checkmark &  &\checkmark \\ \hline
				Pilot-free channel inference  & &   &\checkmark &\checkmark &\checkmark   &\checkmark \\ \hline
				No channel model assumption  &\checkmark &\checkmark  & & &    &\checkmark \\ \hline
				Machine-learning-based approach  &  & \checkmark & &\checkmark &\checkmark  &\checkmark \\ \hline
				Robustness to dynamic environment   &\checkmark  & \checkmark & & & \checkmark &\checkmark  \\ \hline		
				\textbf{Multimodal-sensing-based CSI estimation}  &  &  & &  &    &\checkmark \\ \hline
				\textbf{Flow-matching-based channel inference}  &   &  &   &  &     & \checkmark\\ \hline
				\textbf{Real-time and low-latency requirement}  &   &  &   &  &    & \checkmark\\ \hline
			\end{tabular}
			\label{tab:survey}
		\end{center}
	\end{table*}

    \subsubsection{Environment-Aware Channel Inference} 
    Environmental sensing information has been increasingly exploited to enhance the understanding of wireless channels. Existing studies can be broadly categorized according to their underlying channel modeling assumptions. The first category assumes a statistical channel model and estimates its key parameters such as the path loss \cite{wang2024multi}, Rician K-factor \cite{zhang2025vision}, and root-mean-square delay spread \cite{10720899} from multimodal sensing data, including GPS coordinates, RGB images, LiDAR point clouds, and radar signals. Instead of focusing on long-term statistical models, \cite{9473871} assumes a quasi-static propagation environment, where the channel characteristics are inferred from spatial information using a K-nearest neighbors (KNN) approach that maps user locations to propagation path knowledge for reconstructing the MIMO channel coefficient matrix. Beyond conventional statistical or ray-tracing models, NeRF2 \cite{10.1145/3570361.3592527} learns a neural representation of radio propagation, effectively introducing a novel data-driven propagation mechanism that implicitly captures reflection, diffraction, and scattering effects to predict channel responses. While the above methods are capable of advanced environment-aware channel inference, most of them are model-based and typically assume static environments or rely on single-modality inputs. By contrast, our work adopts a fully data-driven framework that infers complete CSI directly from multimodal sensing data, facilitating robust inference under dynamic wireless environments.  

    \subsubsection{Advances in Flow Matching for Distribution Mapping} 
    Recently, flow matching \cite{lipman2023flow,tong2024improving} has emerged as a promising alternative to diffusion-based \cite{10812969} and GAN-based \cite{10417075} generative models. In contrast to diffusion models that rely on fixed Gaussian priors, flow matching provides a mathematically principled framework for learning continuous mappings between arbitrary distributions, covering both intramodal \cite{liu2023flow, zhou2024denoising, 10.5555/3618408.3619323} and cross-modal \cite{yang2025texttoimage, luo2025curveflow, liu2024flowing, he2025flowtok} scenarios. In intra-modal settings (e.g., face-to-face \cite{liu2023flow, zhou2024denoising} and sketches-to-images \cite{10.5555/3618408.3619323} translations), flow matching learns a continuous transport process that maps one distribution to another. Extending this framework to cross-modal generation (e.g., text-to-image) is more challenging because the source and target reside in distinct data spaces. Early studies adopt Gaussian priors and condition the flow on auxiliary modality features (e.g., text embeddings) to guide this mapping \cite{yang2025texttoimage, luo2025curveflow}, which increases model complexity and often limits generative quality. To overcome these limitations, both CrossFlow \cite{liu2024flowing} and FlowTok \cite{he2025flowtok} project heterogeneous modalities into shared latent spaces, enabling more direct and efficient cross-modal distribution transformation. Despite these advances in computer vision, the potential of flow matching for supporting cross-modal distribution mapping in wireless communications, particularly for wireless channel inference, remains unexplored.

	Our contributions are boldly and explicitly contrasted to the existing literature in Table \ref{tab:survey} and are further described next.		

	\subsection{Contributions and Organization}
    This work investigates environment-aware channel inference in dynamic wireless environments, where we infer complete CSI directly from multimodal sensing data, including camera images, LiDAR point clouds, and GPS coordinates, without relying on pilot signaling. Building on recent advances in flow-based generative modeling, we develop a cross-modal flow framework that bridges the distributions of multimodal sensing data and channel representations to enable real-time and model-free CSI estimation. In contrast to prior sensing-enabled communication approaches tailored to specific tasks such as beamforming, our framework consequently generalizes to downstream physical-layer operations and provides a unified foundation for multimodal sensing and communication integration. The main contributions are summarized below.

	\begin{figure*}[t]
		\centering
		\includegraphics[width = 1\linewidth]{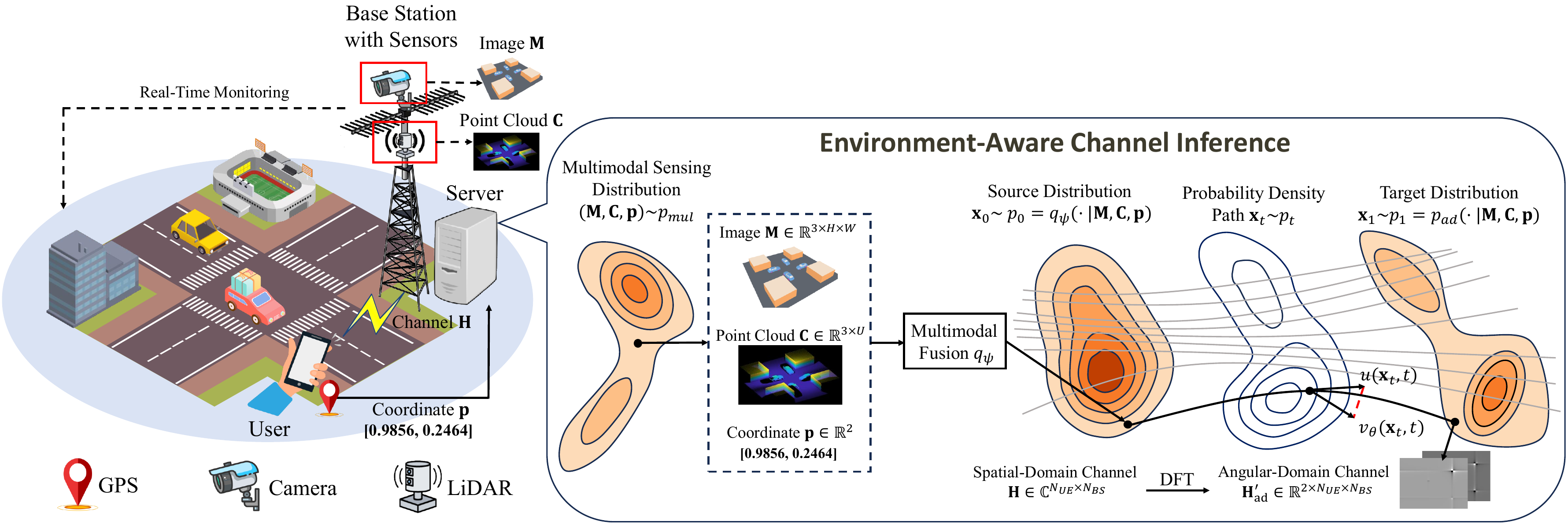}
		\caption{Physical scenario where the base station is equipped with a camera, a LiDAR system and a server, and the user is deployed with a GPS. At the server, the complete CSI between the base station and the user is estimated from multimodal sensing data, including image, point cloud and coordinate.}
		\label{fig:Physical_Scenario}
	\end{figure*}

    \noindent $\bullet$\textbf{Modeling environment-aware channel inference via cross-modal flow:} We model environment-aware channel inference as a cross-modal flow matching problem \cite{liu2024flowing,he2025flowtok}, where the transformation from multimodal sensing distribution to wireless channel distribution is represented as a continuous cross-modal flow. To realize this flow, heterogeneous sensing modalities are fused through a stochastic encoder into a multimodal latent distribution in the channel space. Then, the model learns a time-dependent velocity field that drives the evolution from the fused sensing modalities toward the target channel modality. This formulation serves as the basis for the subsequent learning and inference framework.

    \noindent $\bullet$\textbf{Tractable learning and efficient inference design:} To make the proposed formulation tractable, we reformulate it as a conditional flow matching problem that learns the velocity field conditioned on the accessible samples drawn from the encoded multimodal latent (source) and wireless channel (target) distributions. To reduce the mismatch between the two distributions, we introduce a modality alignment loss that not only regularizes the encoded source distribution but encourages it to closely align with the target channel distribution. During inference, a second-order numerical integration scheme is employed to approximate the learned cross-modal flow for reducing latency and preserve estimation quality.

    \noindent $\bullet$\textbf{Experiments:} We construct a procedural data generator based on Sionna \cite{hoydis2023} and Blender \cite{blender2018} to enable the joint modeling of realistic sensing environments and wireless propagation. System-level experiments show that the proposed framework achieves sub-10 dB normalized mean square error (NMSE) in channel estimation and up to 25\% improvement in spectral efficiency for beamforming as a downstream task, outperforming both pilot-based and sensing-based baselines, while generalizing beyond task-specific multimodal beamforming.

    The remainder of the paper is organized as follows. Section II introduces the system model. Section III presents our problem formulation for environment-aware channel inference. The learning and optimization framework is presented in Section IV. Section V provides experimental results, followed by our conclusions in Section VI.

    \section{System Model}
	\subsection{Physical Scenario and Transmission Protocol}
	As shown in Fig. \ref{fig:Physical_Scenario}, we consider a massive MIMO system operating in the millimeter wave (mmWave) band, where a base station (BS) having $N_{BS}$ transmit antennas (TAs) serves a single user equipped with $N_{UE}$ receive antennas (RAs). The coverage region includes both static and dynamic entities, such as buildings and moving vehicles, which affect the propagation of electromagnetic signals. To enable real-time monitoring of the wireless environment, the BS is equipped with a LiDAR sensor, a panoramic camera, and it also receives GPS information reported by the associated user. Specifically, the LiDAR point cloud, consisting of $U$ three-dimensional points, is denoted by $\mathbf{C}\in\mathbb{R}^{3\times U}$, which characterizes the spatial structure of the environment, including the shape, size, and distance of surrounding objects. The panoramic camera image having a height of $H$, a width of $W$, and RGB channels is denoted by $\mathbf{M}\in \mathbb{R}^{3\times H\times W}$, which captures a 360-degree view of the environment. The two-dimensional GPS coordinate is denoted by $\mathbf{p}\in\mathbb{R}^{2}$, representing the position relative to the BS. All collected multimodal sensing data are stored on a server deployed at the BS, where they are used to infer the complete CSI between the BS and the user. This task is referred to as \textit{environment-aware channel inference}.

	\begin{figure}[t]
		\centering
		\includegraphics[width = 1\linewidth]{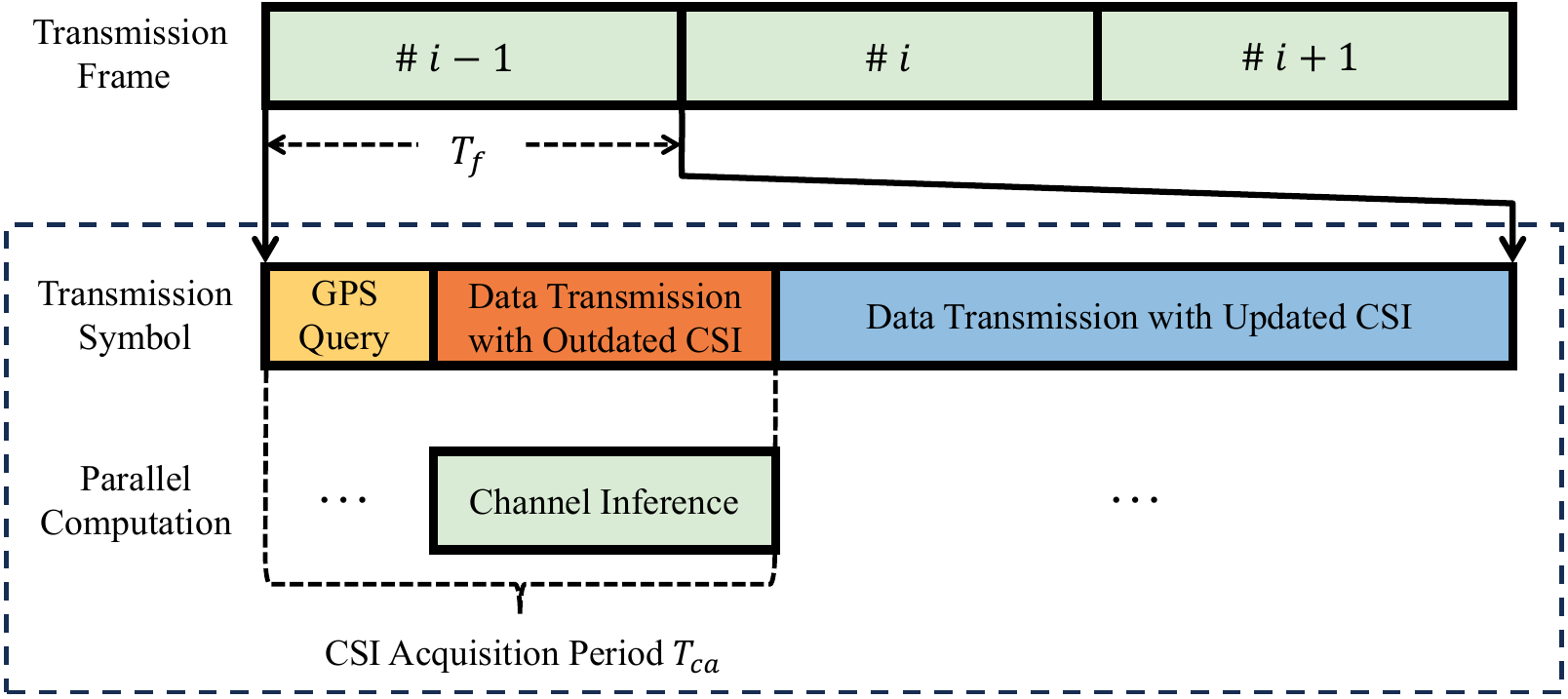}
		\caption{Transmission protocol with frame-based CSI acquisition, where the environment-aware channel inference is employed for CSI update.}
		\label{fig:Frame_Stucture}
	\end{figure}
    The transmission protocol is illustrated in Fig. \ref{fig:Frame_Stucture}, where environment-aware channel inference is employed for CSI acquisition. The transmission resources are divided into consecutive frames, each with a duration of $T_f$. We assume a block-fading channel model in which the CSI remains constant within each frame but varies randomly and independently across frames. Each transmission frame starts with a CSI acquisition period $T_{ca}$, during which the user reports its instantaneous coordinate to the BS in the GPS query phase. This phase is followed by a channel inference stage, when the BS infers the CSI of the current frame by exploiting the reported position together with the camera image and LiDAR point cloud, while data transmission continues concurrently using the outdated CSI from the previous frame. Once the inference is completed, the updated CSI is applied for the rest of the frame.

	The proposed environment-aware channel inference framework is positioned as an initial feasibility study conducted under controlled yet representative assumptions. Specifically, we consider multimodal sensing inputs synchronized with channel dynamics, centralized sensing and inference at the BS, as well as site-specific training and evaluation. These assumptions allow us to focus on the potential benefits of cross-modal CSI inference, while the present study does not yet address several more general and practical scenarios, including mismatch between low-rate sensor updates and fast channel variations, distributed multi-view sensing and collaborative inference, or zero-shot generalization to unseen environments. 
    
	\subsection{Channel Representation and Pre/Post-Processing}
    We denote the mmWave MIMO channel between the BS and the user by $\mathbf{H} \in \mathbb{C}^{N_{UE} \times N_{BS}}$. Due to the limited number of propagation clusters and the narrow angular spread, the mmWave channel exhibits sparsity in the angular domain. Motivated by this observation and following \cite{10705115}, we perform channel inference in the angular domain. To this end, the spatial-domain channel matrix $\mathbf{H}$ is transformed into its angular-domain counterpart $\mathbf{H}_{\text{ad}}$ through discrete Fourier transforms (DFT) as   
    \begin{align}
   	    	&\mathbf{H}_{\text{ad}}= \mathbf{F}_{N_{UE}}^{H}\mathbf{H}\mathbf{F}_{N_{BS}},
   	\end{align}
    where   $\mathbf{F}_{N_{UE}}\in \mathbb{C}^{N_{UE} \times N_{UE}}$ and $\mathbf{F}_{N_{BS}}\in \mathbb{C}^{N_{BS} \times N_{BS}}$  represent the DFT matrices defined as
    \begin{align}
    	&[\mathbf{F}_{N_{UE}}]_{m, k} = \frac{1}{\sqrt{{N_{UE}}}} e^{j \tfrac{2\pi}{{N_{UE}}} (m-1)(k-1)}, \quad m,k = 1,\ldots,{N_{UE}},\\
        &[\mathbf{F}_{N_{BS}}]_{m, k} = \frac{1}{\sqrt{N_{BS}}} e^{j \tfrac{2\pi}{N_{BS}} (m-1)(k-1)}, \quad m,k = 1,\ldots,{N_{BS}}.
    \end{align}

    Since conventional neural networks operate on real-valued tensors, the complex angular-domain channel matrix $\mathbf{H}_{\text{ad}}$ is converted into a real-valued representation before being fed into the learning model. Specifically, we stack its real and imaginary components along the first dimension to form $\mathbf{H}^{\prime}_{\text{ad}} \in \mathbb{R}^{2 \times N_{UE}\times N_{BS}}$. After inference, the resultant real-valued tensor $\widehat{\mathbf{H}}'_{\text{ad}}$ is transformed back into its complex form as 
	\begin{align}\label{eq:back_to_complex}
	   	 \widehat{\mathbf{H}}_{\text{ad}} = \widehat{\mathbf{H}}^{\prime}_{\text{ad}}(1,:,:) + j \, \widehat{\mathbf{H}}^{\prime}_{\text{ad}}(2,:,:), 
	\end{align} 
    yielding the estimated angular-domain channel matrix. Then, the estimated channel $\widehat{\mathbf{H}}$ in the spatial domain is obtained by the inverse DFT operation
    \begin{align}\label{eq:back_to_sptial}
    	&\widehat{\mathbf{H}}= \mathbf{F}_{N_{UE}}\widehat{\mathbf{H}}_{\text{ad}}\mathbf{F}_{N_{BS}}^{H}. 
    \end{align}

    \section{Problem Formulation for Environment-Aware \\ Channel Inference}
    As established in channel modeling studies, the wireless channel between a transmitter and a receiver is primarily determined by their relative geometry and the surrounding environment \cite{10599118}. This intrinsic dependency provides a physical basis for inferring high-dimensional channel matrices from multimodal environmental sensing data. Motivated by this observation, we seek to model the multimodal sensing-conditioned channel distribution $p_{ad}(\mathbf{H}'_{\text{ad}}\mid \mathbf{M}, \mathbf{C}, \mathbf{p})$ and sample channel realizations from its high-probability regions.

    To this end, we view both the multimodal environmental sensing data and the wireless channel as random variables that follow different probability distributions. Specifically, the sensing distribution characterizes the stochastic variations in LiDAR, camera, and the GPS data gleaned from different environmental configurations, while the channel distribution represents the corresponding variations in the channel matrices. This distributional formulation naturally fits within the flow matching framework \cite{lipman2023flow}, which models data generation as a continuous transformation from a source to a target distribution guided by a time-dependent velocity field.
    
    However, in environment-aware channel inference, the source and target distributions correspond to heterogeneous modalities lying in different data spaces, which makes standard flow matching formulations inapplicable \cite{lipman2023flow,tong2024improving}. To generalize flow matching to this setting, we propose a cross-modal flow framework that first fuses heterogeneous sensing modalities into a shape-aligned channel space to construct a latent distribution, and then learns a distribution-level mapping from the latent distribution to the target channel distribution. This framework is conceptually related to the CrossFlow \cite{liu2024flowing}, which adapts flow matching to a cross-domain generation task, but differs fundamentally in task setting, problem formulation, and domain-specific design. While CrossFlow is designed for cross-modal media generation tasks such as text-to-image generation, image captioning, and depth estimation, our method targets full CSI inference from multimodal sensing data. This leads to a many-to-one sensing-to-channel generation problem rather than a one-to-one modality translation problem, which requires multimodal fusion and cross-modal alignment before establishing the cross-modal mapping. In addition, our framework calls for wireless-specific design beyond CrossFlow, including a sparse angular-domain representation tailored to wireless channels and a low-latency inference scheme for real-time CSI acquisition.

    As illustrated in Fig. \ref{fig:Physical_Scenario}, we formalize the environment-aware channel inference problem by considering multimodal sensing-channel observations $(\mathbf{M}, \mathbf{C}, \mathbf{p}, \mathbf{H}'_{\text{ad}})$ jointly drawn from an environmental distribution 
	\begin{align} \label{eq:data distribution}p_{\text{env}}(\mathbf{M}, \mathbf{C}, \mathbf{p}, \mathbf{H}'_{\text{ad}})=p_{mul}(\mathbf{M}, \mathbf{C}, \mathbf{p}) p_{ad}(\mathbf{H}'_{\text{ad}}\mid \mathbf{M}, \mathbf{C}, \mathbf{p}), 
	\end{align}
	where $p_{mul}(\mathbf{M}, \mathbf{C}, \mathbf{p})$ denotes the joint distribution of multimodal sensing observations. The camera RGB image $\mathbf{M}$, LiDAR point cloud $\mathbf{C}$, and GPS coordinate $\mathbf{p}$ are fused through a stochastic encoder $q_\psi$ having learnable parameters $\psi$, yielding a conditional latent distribution 
	$q_\psi(\cdot\mid\mathbf{M}, \mathbf{C}, \mathbf{p})$ that serves as the source distribution $p_0$.  Conditioned on the same multimodal sensing observations, the conditional channel distribution $p_{\text{ad}}( \cdot \mid \mathbf{M}, \mathbf{C}, \mathbf{p})$ functions as the target distribution $p_1$. The continuous transformation from the source distribution $p_0$ to the target distribution $p_1$ is described by a probability path $\{p_t\}_{t\in[0,1]}$ that smoothly interpolates between them. This path is governed by a time-dependent velocity field $u(\mathbf{x}_t,t)$, so that the evolution of an intermediate state $\mathbf{x}_t \sim p_t$ follows the ordinary differential equation (ODE)
	\begin{align}
		\frac{d\mathbf{x}_t}{dt} = u(\mathbf{x}_t,t),
		\label{eq:ODE_definition_a}
	\end{align}
	which characterizes the instantaneous flow that is capable of transporting samples from the multimodal latent distribution $q_\psi$ 
	toward the channel distribution $p_{\text{ad}}$. Since the true velocity field $u(\mathbf{x}_t,t)$ implied by the probability path is generally intractable, 
	we approximate it using a neural network $v_\theta(\mathbf{x}_t,t)$ relying on the learnable parameters $\theta$. Following the standard flow matching formulation in \cite{lipman2023flow}, the learning objective is to minimize the expected discrepancy between the neural and true velocity fields:
	\begin{align}
		\text{(P1)}~~
		\min_{\psi,\theta}~ \mathcal{J}_{\mathrm{FM}} 
		= \mathbb{E}_{t \sim \mathcal{U}[0,1],\, \mathbf{x}_t \sim p_t}
		\bigl\lVert v_\theta(\mathbf{x}_t,t) - u(\mathbf{x}_t,t) \bigr\rVert_F^2.
		\label{eq:FlowMatching}
	\end{align} 

	\section{Learning and Optimization for Environment-Aware \\ Channel Inference}
    In this section, we propose a cross-modal flow framework to realize environment-aware channel inference. Specifically, we introduce the learning objectives for cross-modal flow in Sec.~IV-A, describe the model architecture and parameterization in Sec.~IV-B, elaborate on the optimization and inference pipeline in Sec.~IV-C, and present the computational complexity analysis in Sec.~IV-D.
    
    \subsection{Learning Objectives for Cross-Modal Flow}
    
    To achieve environment-aware channel inference, we design learning objectives for tractable cross-modal flow. Specifically, we reformulate the original flow matching objective in \eqref{eq:FlowMatching} to enable Monte Carlo approximation, and introduce an extra modality alignment loss for supporting the evolution from multimodal sensing data toward the channel representation. Finally, the overall learning objective is proposed. 

    \subsubsection{Flow Matching Loss Reformulation}
    Although the flow matching objective in \eqref{eq:FlowMatching} provides a theoretically elegant formulation, it is intractable to use in practice. This is because the true probability path $p_t$ and its corresponding velocity field $u(\mathbf{x}_t,t)$ are unknown and cannot be explicitly constructed without access to the underlying generative process that transforms $p_0$ to $p_1$. In other words, during model training, we can only observe the initial state $\mathbf{x}_0$ and the terminal state $\mathbf{x}_1$ from the source latent distribution $q_\psi$ (i.e., $p_0$) and the target channel distribution $p_{\text{ad}}$ (i.e., $p_1$), respectively, but we have no prior knowledge of an appropriate pair $[p_t, u(\mathbf{x}_t,t)]$ that yields a continuous flow to connect them. To address this issue, we construct a conditional probability path $p_t(\mathbf{x}_t ; \mathbf{x}_0, \mathbf{x}_1)$ and a corresponding conditional velocity field $u(\mathbf{x}_t, t; \mathbf{x}_0, \mathbf{x}_1)$, which describe the evolution of intermediate states $\mathbf{x}_t$ conditioned on the accessible paired sensing-channel representations $(\mathbf{x}_0, \mathbf{x}_1)$ drawn from a joint distribution, formally defined as
    \begin{align}
    (\mathbf{x}_0, \mathbf{x}_1)\sim \pi(\mathbf{x}_0, \mathbf{x}_1), 
    \end{align} 
    whose marginal distributions are  $q_\psi$ and $p_\text{ad}$, respectively.  
    
    To make the conditional path $p_t(\mathbf{x}_t ; \mathbf{x}_0, \mathbf{x}_1)$  analytically tractable, we follow \cite{tong2024improving} and model it as an element-wise Gaussian distribution with mean $t\mathbf{x}_1 + (1-t)\mathbf{x}_0$ and isotropic variance $\sigma_\text{min}^2$, implying element-wise independence within $\mathbf{x}_t$. The marginal probability path can be obtained by integrating out the joint distribution of the initial and terminal states:
    \begin{equation}\label{eq: marginal prob path}
    p_t(\mathbf{x}_t)
    = \int p_t(\mathbf{x}_t;\mathbf{x}_0,\mathbf{x}_1)\,
        \pi(\mathbf{x}_0,\mathbf{x}_1)\,
        d\mathbf{x}_0\,d\mathbf{x}_1.
    \end{equation}
    Given the definition of $p_t(\mathbf{x}_t;\mathbf{x}_0,\mathbf{x}_1)$, its mean reduces to $\mathbf{x}_0$ and $\mathbf{x}_1$ at $t=0$ and $t=1$, respectively. Substituting these endpoint cases into \eqref{eq: marginal prob path} shows that 
    the marginal distributions at the boundaries correspond to Gaussian-smoothing of the source and target distributions, i.e.,
    \begin{equation}
    p_0 = q_\psi * \mathcal{N}_{\mathrm{iid}}(0, \sigma_{\min}^2), 
    \qquad
    p_1 = p_{\mathrm{ad}} * \mathcal{N}_{\mathrm{iid}}(0, \sigma_{\min}^2),
    \end{equation}
    where $*$ denotes convolution over the tensor domain. As $\sigma_{\min} \rightarrow 0$, the marginal distributions converge to $p_0 = q_{\psi}$ and $p_1 = p_{\text{ad}}$, which are the desired source and target distributions defined in the problem formulation.
    
    Having established the conditional probability path that satisfies the desired marginal conditions, we next derive the corresponding conditional velocity field through the ODE in (\ref{eq:ODE_definition_a}). According to the definition of conditional Gaussian path, the intermediate state $\mathbf{x}_t \sim p_t(\mathbf{x}_t; \mathbf{x}_0, \mathbf{x}_1)$ 
    can be represented as
    \begin{equation}\label{eq:Intermediate_Variable}
        \mathbf{x}_t =  t\mathbf{x}_1 + (1-t)\mathbf{x}_0 + \sigma_{\min}\boldsymbol{\epsilon},
    \end{equation}
    where the elements of $\boldsymbol{\epsilon}$ are independently and identically distributed (i.i.d.) as $\mathcal{N}(0,1)$. Taking the time derivative of (\ref{eq:Intermediate_Variable}) yields the conditional velocity field as follows:
    \begin{align}\label{eq:conditional_velocity_field}
        u(\mathbf{x}_t,t ; \mathbf{x}_0, \mathbf{x}_1) 
        &= \frac{d}{dt}\Big((1-t)\mathbf{x}_0 + t\mathbf{x}_1 + \sigma_{\min}\boldsymbol{\epsilon}\Big)   \\
        &= \mathbf{x}_1 - \mathbf{x}_0. \notag
    \end{align}

    Given the conditional probability path $p_t(\mathbf{x}_t ; \mathbf{x}_0, \mathbf{x}_1)$ and velocity field $u(\mathbf{x}_t,t;\mathbf{x}_0, \mathbf{x}_1)$,  we extend the original flow matching objective in (\ref{eq:FlowMatching}) by replacing the probability path and velocity field with their conditional counterparts and introducing an expectation over the joint distribution $\pi(\mathbf{x}_0,\mathbf{x}_1)$. The resultant conditional flow matching (CFM) objective is given as follows: 
    \begin{align}\label{eq:CFM_objective}
        &\mathcal{J}_{\mathrm{CFM}}= \mathbb{E}_{t \sim \mathcal{U}[0,1],\,(\mathbf{x}_0, \mathbf{x}_1)\sim \pi,\, \mathbf{x}_t \sim p_t(\cdot;\mathbf{x}_0, \mathbf{x}_1)} \bigl\lVert v_\theta(\mathbf{x}_t,t) - u(\mathbf{x}_t,t;\mathbf{x}_0, \mathbf{x}_1)\bigr\rVert_F^2, \notag\\
    	&~~\quad\overset{\sigma_{min}\rightarrow0}{=} \mathbb{E}_{t \sim \mathcal{U}[0,1],\,(\mathbf{x}_0,\mathbf{x}_1)\sim\pi}
        \bigl\lVert v_\theta\Big(t\mathbf{x}_1+ (1-t)\mathbf{x}_0,t\Big)
        - (\mathbf{x}_1-\mathbf{x}_0)\bigr\rVert_F^2, 
    \end{align}
    where the second equality follows from the representation of $\mathbf{x}_t$ in \eqref{eq:Intermediate_Variable} as $\sigma_\text{min} \rightarrow 0$. It has been shown in~\cite{lipman2023flow,tong2024improving} that the conditional formulation in (\ref{eq:CFM_objective}) achieves the same gradient dynamics as the original flow matching objective in (\ref{eq:FlowMatching}), i.e., $\nabla \mathcal{J}_{\mathrm{FM}} = \nabla \mathcal{J}_{\mathrm{CFM}}$. This indicates that CFM provides a tractable yet theoretically equivalent alternative to the intractable original formulation, which enables efficient optimization of $\theta$ and $\psi$ via (\ref{eq:CFM_objective}).

    In particular, we model the latent distribution $q_{\psi}(\mathbf{x}_0 \mid \mathbf{M}, \mathbf{C}, \mathbf{p})$ 
    as a multivariate Gaussian distribution with mean  $\boldsymbol{\mu}(\mathbf{M}, \mathbf{C}, \mathbf{p}; \psi) \in \mathbb{R}^{2 \times N_{UE}\times N_{BS}}$  and standard deviation $\boldsymbol{\sigma}(\mathbf{M}, \mathbf{C}, \mathbf{p}; \psi) \in \mathbb{R}_{\geq 0}^{2 \times N_{UE}\times N_{BS}}$, where both $\boldsymbol{\mu}$ and $\boldsymbol{\sigma}$ are outputs of a neural network parameterized by $\psi$ with inputs $(\mathbf{M}, \mathbf{C}, \mathbf{p})$. Using the reparameterization technique, the initial state is sampled as
    \begin{align}\label{eq:VE_repara}
        \mathbf{x}_0 = 
        \boldsymbol{\mu}(\mathbf{M}, \mathbf{C}, \mathbf{p}; \psi)
        + \boldsymbol{\sigma}(\mathbf{M}, \mathbf{C}, \mathbf{p}; \psi) \odot \boldsymbol{\epsilon},
    \end{align}
    where the elements of $\boldsymbol{\epsilon}$ are  i.i.d. as $\mathcal{N}(0,1)$ and $\odot$ denotes element-wise multiplication. Given a batch of paired multimodal sensing-channel samples $\{\mathbf{M}^{[s]}, \mathbf{C}^{[s]}, \mathbf{p}^{[s]}, \mathbf{H}_{\text{ad}}'^{[s]}\}_{s=1}^{S}$ and time indices $\{t^{[s]}\}_{s=1}^{S}$ independently sampled from the uniform distribution $\mathcal{U}[0,1]$, we approximate the expectation in \eqref{eq:CFM_objective} via Monte Carlo sampling, and thus, obtain the empirical CFM loss:
    \begin{align}\label{eq:CFM_Loss}
        \mathcal{L}_{\mathrm{CFM}} 
        = \frac{1}{S}\sum_{s=1}^{S}
        \bigl\lVert 
            v_{\theta}\bigl(t^{[s]}\mathbf{H}_{\text{ad}}^{\prime[s]}+(1-t^{[s]})\mathbf{x}_0^{[s]}, t^{[s]}\bigr)
            - \bigl(\mathbf{H}_{\text{ad}}^{\prime[s]} - \mathbf{x}_0^{[s]}\bigr)\bigr\rVert_F^2.
    \end{align}

    \begin{figure}[t]
    	\centering
    	\includegraphics[width = 1\linewidth]{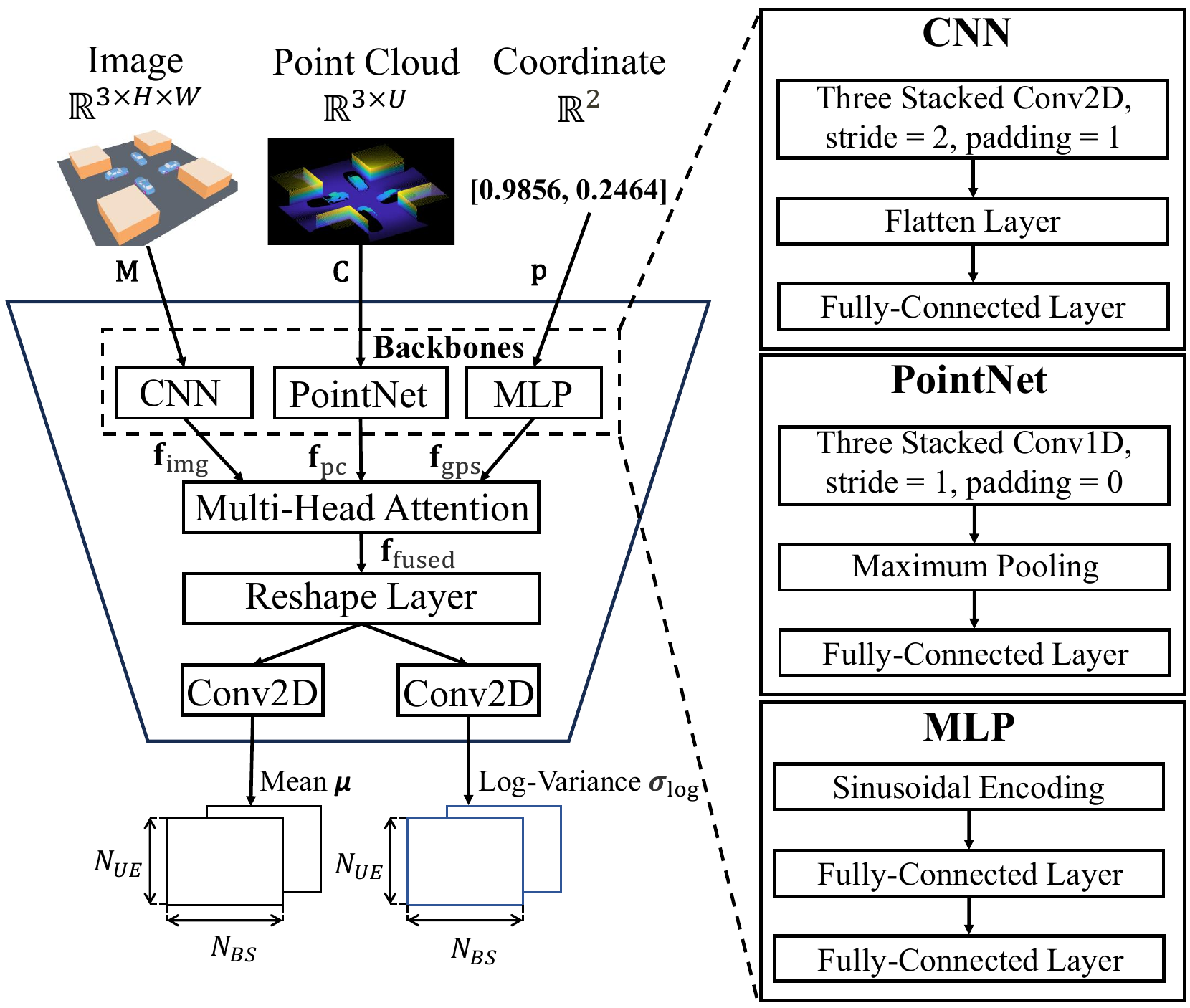}
    	\caption{Network structure of the multimodal stochastic encoder.}
    	\label{fig:Variational_Encoder}
    \end{figure}

    \subsubsection{Modality Alignment Loss}   
    In the cross-modal flow, training the stochastic encoder solely with the conditional flow matching loss may lead to weak alignment between source multimodal representations and target channel samples, which can make the induced probability path unnecessarily irregular and finally degrade the generative performance. To improve transport efficiency, we adopt a mutual-information-inspired contrastive learning strategy \cite{radford2021learning}, which encourages the paired multimodal latent representations and channel samples to have higher similarity than the unpaired ones. However, we emphasize that this strategy does not constitute a formal derivation of the transport path with minimal cost, but instead serves as an empirical regularizer that facilitates smoother and more stable flow matching.

	\begin{figure}[t]
		\centering
		\includegraphics[width = 1\linewidth]{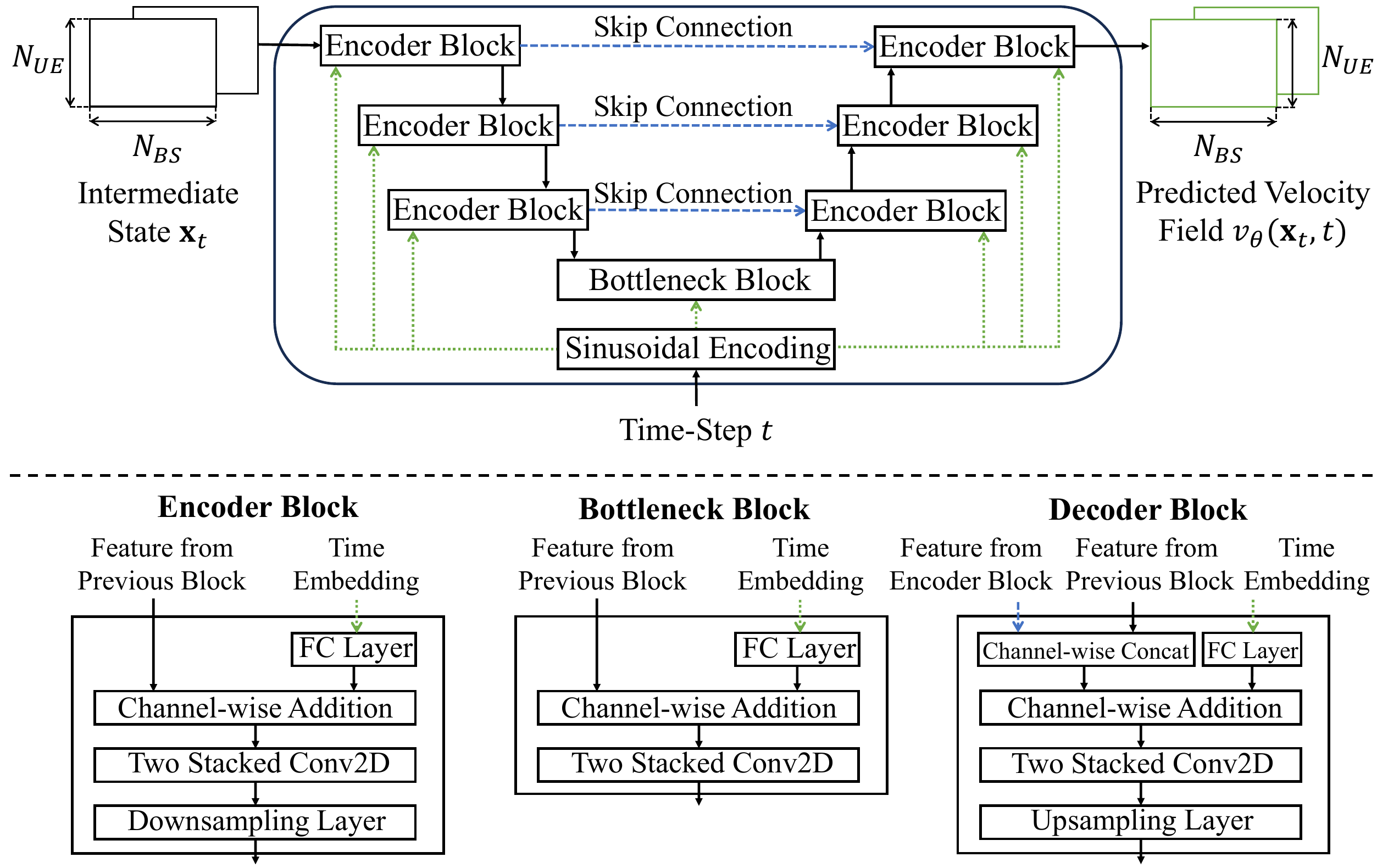}
		\caption{Network structure of the neural velocity field.}
		\label{fig:U-Net}
	\end{figure}

    Accordingly, we design the \textit{modality alignment loss} that jointly minimizes the contrastive loss to enhance cross-modal consistency and incorporates a KL-divergence term to regularize the encoded source distribution:
    \begin{align}\label{eq:MA_Loss}
        \mathcal{L}_{\mathrm{MA}}
        = -\frac{1}{S}\sum_{s=1}^{S}
            \log
            \frac{
                \exp\big(\text{sim}(\mathbf{x}_0^{[s]},\mathbf{H}_{\text{ad}}^{\prime[s]})/\tau\big)
            }{
                \sum_{k=1}^{S}\exp\big(\text{sim}(\mathbf{x}_0^{[s]},\mathbf{H}_{\text{ad}}^{\prime[k]})/\tau\big)
            }
            + \lambda    \mathcal{L}_{\mathrm{KL}},
    \end{align}
    where $\tau$ denotes the trainable temperature coefficient controlling the contrastive sharpness,  $\lambda$ is the hyperparameter that balances the alignment and regularization terms, and  $\text{sim}(\cdot,\cdot)$ represents the cosine similarity, while the KL-divergence loss is defined as
    \begin{align}\label{eq:KL_Loss}
        &\mathcal{L}_{\mathrm{KL}} = -\frac{1}{S}\sum_{s=1}^{S}  \boldsymbol{1}^T  \text{vec}\Big(1 + \log\Big(\boldsymbol{\sigma}^2(\mathbf{M}^{[s]}, \mathbf{C}^{[s]}, \mathbf{p}^{[s]}; \psi)\Big) \\
        &\qquad\qquad - \boldsymbol{\mu}^2(\mathbf{M}^{[s]}, \mathbf{C}^{[s]}, \mathbf{p}^{[s]}; \psi) - \boldsymbol{\sigma}^2(\mathbf{M}^{[s]}, \mathbf{C}^{[s]}, \mathbf{p}^{[s]}; \psi) \Big).\notag
    \end{align} 

    \subsubsection{Overall Learning Objective} 
    The overall learning objective jointly optimizes the CFM loss $\mathcal{L}_{\mathrm{CFM}}$ in (\ref{eq:CFM_Loss}) and the modality alignment loss $\mathcal{L}_{\mathrm{MA}}$ in (\ref{eq:MA_Loss}) to achieve robust and efficient sensing-to-channel evolution. The associated optimization problem is formulated as
    \begin{align}\label{eq:overall_objective}
        \min_{\psi,\theta,\tau} \big(\mathcal{L}_{\mathrm{CFM}} + \mathcal{L}_{\mathrm{MA}}\big),
    \end{align}
    which enables joint end-to-end optimization of the multimodal encoder $q_\psi$ and the neural velocity field $v_\theta$.

    \subsection{Model Architecture and Parameterization}
    This section details the network architectures that implement the overall learning objective in \eqref{eq:overall_objective}, which comprise the multimodal stochastic encoder $q_\psi$ that fuses heterogeneous sensing modalities into a latent distribution in channel space, and the neural velocity field $v_\theta$ for flow matching. 

	\begin{figure*}[t]
		\centering
		\includegraphics[width = 0.83\linewidth]{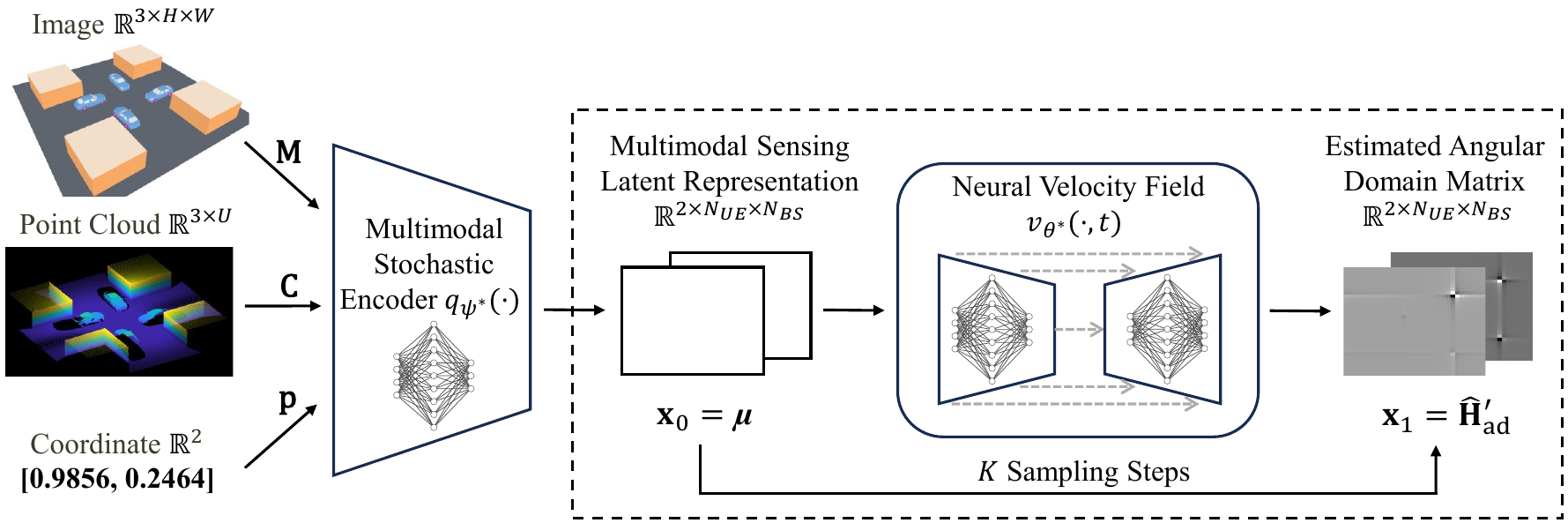}
		\caption{Inference pipeline that enables the cross-modality evolution from multimodal sensing data to wireless channel representation.}
		\label{fig:Channel_Inference_Framework}
	\end{figure*}

    \subsubsection{Multimodal Stochastic Encoder}
    The stochastic encoder $q_{\psi}(\cdot)$ employs modality-specific backbone networks to extract multimodal sensing features, and then fuses them into a latent distribution within the channel space. 

    As detailed in Fig. \ref{fig:Variational_Encoder}, the encoder integrates three backbone networks -- a convolutional neural network (CNN), a point cloud network (PointNet), and a multilayer perceptron (MLP) -- to extract features from the image, point cloud, and coordinates, respectively. This modality-specific design is motivated by the fact that the three input modalities exhibit fundamentally different structural properties. Using a single network to process all modalities would force a shared feature-extraction strategy to accommodate heterogeneous inputs, causing the learned features to become a compromise rather than modality-specific optimal representations. Specifically, \textbf{(i) the CNN architecture} consists of three stacked 2D convolutional (Conv2D) layers with progressively doubled channel sizes, where each layer halves the height and width of the feature map via a stride-2 downsampling operation. The resultant feature map from the final Conv2D layer is flattened into a vector, and then processed by a fully-connected (FC) layer to produce a feature vector $\mathbf{f}_{\text{img}}\in \mathbb{R}^{d}$; \textbf{(ii) the PointNet architecture} applies three stacked 1D convolutional (Conv1D) layers to expand the dimensionality of each 3D point, followed by a max-pooling operation across the point dimension and a FC layer to obtain a global feature vector $\mathbf{f}_{\text{pc}}\in \mathbb{R}^{d}$; \textbf{(iii) the MLP architecture} applies a sinusoidal encoding module \cite{10.5555/3295222.3295349} to transform the coordinate into a high-dimensional position embedding, followed by two sequential FC layers that generate the corresponding feature vector $\mathbf{f}_{\text{gps}}\in \mathbb{R}^{d}$. 
    
    The resulting modality-specific feature vectors $\mathbf{f}_{\mathrm{img}}$, $ \mathbf{f}_{\mathrm{pc}}$, $\mathbf{f}_{\mathrm{gps}}$, each of dimension $d$, are fused through a multi-head attention layer \cite{10.5555/3295222.3295349}. The rationale for this design is that the attention mechanism enables each modality to selectively aggregate complementary information from the others while reducing the impact of irrelevant or noisy features, and the multiple heads capture diverse cross-modal correlations in parallel subspaces to improve the expressiveness of the fused representation. Specifically, the three feature vectors are stacked along the modality axis and treated as a feature matrix:
    \begin{equation}
    	\mathbf{F} =
    	\begin{bmatrix}
    		\mathbf{f}_{\mathrm{img}},
    		\mathbf{f}_{\mathrm{pc}},
    		\mathbf{f}_{\mathrm{gps}}
    	\end{bmatrix}^\top
    	\in \mathbb{R}^{3\times d}.
    \end{equation}
    For the $m$-th attention head, the stacked feature matrix is linearly projected into the query, key and value spaces as
    \begin{align}
    	\mathbf{Q}_m = \mathbf{F}\mathbf{W}_m^Q \in \mathbb{R}^{3\times d_h}, \\
    	\mathbf{K}_m = \mathbf{F}\mathbf{W}_m^K \in \mathbb{R}^{3\times d_h}, \\
    	\mathbf{V}_m = \mathbf{F}\mathbf{W}_m^V \in \mathbb{R}^{3\times d_h},
    \end{align}
    where $\mathbf{W}_m^Q, \mathbf{W}_m^K, \mathbf{W}_m^V \in \mathbb{R}^{d\times d_h}$ are the learnable attention projection matrices, and $d_h$ denotes the dimension of each attention head. The $m$-th head first calculates its attention matrix as
    \begin{equation}
    	\mathbf{A}_m
    	=
    	\operatorname{Softmax}
    	\left(
    	\frac{\mathbf{Q}_m\mathbf{K}_m^\top}{\sqrt{d_h}}
    	\right)
    	\in \mathbb{R}^{3\times3},
    \end{equation}  
    and then performs scaled dot-product attention to produce the attended representation as
	\begin{equation}
		\mathbf{O}_m
		=
		\mathbf{A}_m \mathbf{V}_m
		\in \mathbb{R}^{3\times d_h}.
	\end{equation}     
    The outputs of all $M$ attention heads are concatenated and projected back to the original feature space:
    \begin{equation}
    	\mathbf{O}
    	=
    	\operatorname{Concat}(\mathbf{O}_1,\mathbf{O}_2,\ldots,\mathbf{O}_M)\mathbf{W}^O
    	\in \mathbb{R}^{3\times d},
    \end{equation}
    where $\mathbf{W}^O \in \mathbb{R}^{M d_h \times d}$ is the learnable output projection matrix. The projected output is aggregated by mean pooling along the modality dimension and passed through a ReLU activation to obtain a fused feature:
    \begin{equation}
    	\mathbf{f}_{\mathrm{fused}}
    	=
    	\operatorname{ReLU}
    	\left(
    	\frac{1}{3}\sum_{i=1}^{3}\mathbf{O}[i,:]
    	\right)
    	\in \mathbb{R}^{d}.
    \end{equation}
    In practice, the feature dimension $d$ is chosen as $2 N_{UE} N_{BS}$, which allows the fused feature $\mathbf{f}_{\mathrm{fused}}$ to be naturally reshaped into a dual-channel feature map aligned with the CSI structure. This feature map is then processed by two independent Conv2D layers to produce a mean of $\boldsymbol{\mu}\in\mathbb{R}^{2 \times N_{UE}\times N_{BS}}$ and a log-variance of $\boldsymbol{\sigma}_{\mathrm{log}}\in\mathbb{R}^{2 \times N_{UE}\times N_{BS}}$, respectively. The variance is obtained via $\boldsymbol{\sigma}^2= \exp(\boldsymbol{\sigma}_{\mathrm{log}})$,  which ensures its positivity and thus enhances numerical stability during model training and inference.

    \begin{algorithm}[t]
    	\caption{Training a Cross-Modal Flow for Environment-Aware Channel Inference}\label{alg:train_crossmodal}
    	\normalsize
    	\begin{algorithmic}[1]
    		\REQUIRE~ Training dataset $\mathcal{D}=\{\mathbf{M}^{[s]},\mathbf{C}^{[s]},\mathbf{p}^{[s]},\mathbf{H}^{\prime[s]}_{\text{ad}}\}_{s=1}^{S}$.
    		\STATE Initialize batch size $B_s$, learning rate $\eta$, weight $\lambda$, number of training epochs $N_{tr}$. 
    		\STATE Randomly initialize the multimodal stochastic encoder, the neural velocity field and the temperature coefficient with $\psi^{(0)}$, $\theta^{(0)}$ and $\tau^{(0)}$, respectively.
    		\FOR{$n = 1$ to $N_{tr}$ epochs}
    		\STATE Randomly extract a subset $\mathcal{D}_s\subset\mathcal{D}$ with $B_s$ data samples and independently sample $B_s$ time indices  $\{\,t^{[s]}\}_{s=1}^{B_s}$ from the uniform distribution $\mathcal{U}[0,1]$.
    		\STATE Compute the mini-batch estimate of the loss in~\eqref{eq:overall_objective} using the sampled subset $\mathcal{D}_s$ and time indices $\{t^{[s]}\}$, and obtain the updated parameters $\psi^{(n)}$, $\theta^{(n)}$, and $\tau^{(n)}$ via Adam with learning rate $\eta$.            
    		\ENDFOR	
    		\RETURN Optimized parameters $\psi^{*}=\psi^{(N_{tr})}$ and $\theta^{*}=\theta^{(N_{tr})}$.
    	\end{algorithmic}
    \end{algorithm}

    \subsubsection{Neural Velocity Field}
    The neural velocity field $v_{\theta}(\cdot,t)$ is designed for predicting the current velocity field from the intermediate state $\mathbf{x}_t$. The U-Net architecture of \cite{ronneberger2015u} fits this task well due to its symmetric encoder-decoder design, which preserves spatial dimension between the input and output, while enabling multi-scale feature extraction and flexible conditional information injection. Motivated by these properties, we extend the classic U-Net by incorporating the time step $t$ as an additional conditioning input for flow matching. 

	As illustrated in Fig. \ref{fig:U-Net}, the network consists of an encoding path, a bottleneck block, and a decoding path connected through skip connections. The time step $t$ is first transformed into a $b$-dimensional embedding using a sinusoidal encoding module \cite{10.5555/3295222.3295349} and injected into each block via FC projections. {\bf (i) Encoding path:} The intermediate state $\mathbf{x}_t$ is progressively downsampled through three encoder blocks, where time embeddings are added to feature maps before applying two consecutive Conv2D layers and a downsampling operation that reduces the spatial size by a factor of four. {\bf(ii) Bottleneck block:} The lowest-resolution feature map is fused with the time embedding and processed by two stacked Conv2D layers to capture global context. {\bf (iii) Decoding path:} The feature map is upsampled through three decoder blocks that concatenate skip connections from the corresponding encoder layers, integrate time embeddings, and apply two concatenated Conv2D layers followed by upsampling to restore spatial resolution. The final output $v_{\theta}(\mathbf{x}_t,t)$ represents the predicted velocity field at time step $t$.

    \subsection{Training and Inference Procedures}
    The parameters of the multimodal stochastic encoder $\psi$ and the neural velocity field $\theta$, as well as the temperature coefficient $\tau$ are jointly optimized with the overall learning objective (\ref{eq:overall_objective}). In this paper, we opt for the adaptive moment estimation (Adam) method as the gradient-descent optimizer, because it offers fast and stable convergence by adaptively adjusting per-parameter learning rates. The training process is detailed in Algorithm \ref{alg:train_crossmodal}. 

    With the learned parameters $\psi^*$ and $\theta^*$, Fig.~\ref{fig:Channel_Inference_Framework} shows the pipeline for estimating the wireless channel from multimodal sensing data. During channel inference, the multimodal stochastic encoder $q_{\psi^{*}}$ encodes the RGB image $\mathbf{M}$, the point cloud $\mathbf{C}$, and the position coordinate $\mathbf{p}$ into the mean $\boldsymbol{\mu}$ and the standard deviation $\boldsymbol{\sigma}$. Whilst the initial state is sampled via the reparameterization trick (\ref{eq:VE_repara}) during training to support Monte Carlo optimization of the CFM objective, we use deterministic maximum-a-posteriori (MAP) initialization during inference by setting $\mathbf{x}_0=\boldsymbol{\mu}$. This choice does not alter the generative nature of the proposed cross-modal flow, but serves as a practical inference strategy that fixes the ODE initial state to a representative high-probability point of the learned Gaussian latent distribution, thereby improving the stability and accuracy of multimodal CSI inference. To achieve the terminal state, we integrate the ODE in (\ref{eq:ODE_definition_a}) from $t=0$ to $t=1$, where the true velocity field $u(\mathbf{x}_{t}, t)$ is replaced by the learned neural velocity field $v_{\theta}(\mathbf{x}_{t}, t)$. In practice, the integration interval $t\in [0,1]$ is discretized into $K$ steps with step size $h = 1/K$. Over each step $k$, we have
    \begin{align} \mathbf{x}_{(k+1)h}-\mathbf{x}_{kh} &= \int_{kh}^{kh+h} v_{\theta^{*}}(\mathbf{x}_{t},t) dt,\\ 
        &\approx hv_{\theta^{*}}(\mathbf{x}_{kh},kh), \notag
    \end{align}
    which corresponds to the first-order (forward Euler) update.

	\begin{algorithm}[t]
		\caption{Pipeline for Online Channel Inference}\label{alg:inference_crossmodal}
		\normalsize
		\begin{algorithmic}[1]
			\REQUIRE~ The integration step number $K$, the RGB image $\mathbf{M}$, the point cloud $\mathbf{C}$,  the position coordinate $\mathbf{p}$, and the learned model parameters $\psi^{*}$ and $\theta^{*}$.
			\STATE Feed $\mathbf{M}$, $\mathbf{C}$, and $\mathbf{p}$ into the multimodal stochastic encoder $q_{\psi^{*}}$ to obtain the mean $\boldsymbol{\mu}$, and set the initial state to $\mathbf{x}_0=\boldsymbol{\mu}$.
			\STATE Initialize the intermediate state $\mathbf{x}_{h}$ at $k=0$ via (\ref{eq:Euler_step}). 
			\FOR{$k=1$ to $K-1$}
			\STATE Update the intermediate state $\mathbf{x}_{kh}$ to $\mathbf{x}_{(k+1)h}$ via (\ref{eq:discrete_update}).
			\ENDFOR
			\STATE Extract the terminal state as estimated channel $\mathbf{\widehat{H}}_{\text{ad}}^{\prime} \leftarrow \mathbf{x}_{1}$.
			\STATE Convert $\mathbf{\widehat{H}}_{\text{ad}}^{\prime}$ into complex-valued form $\mathbf{\widehat{H}}_{\text{ad}}$ using Eq. (\ref{eq:back_to_complex}), and then transform it into spatial domain $\mathbf{\widehat{H}}$ via Eq. (\ref{eq:back_to_sptial}).
			\RETURN Estimated  channel coefficient matrix $\widehat{\mathbf{H}}$.
		\end{algorithmic}
	\end{algorithm}

    To improve the numerical accuracy of the integration, we adopt the second-order Adams-Bashforth method \cite{hairer1993solving}. The update for $k = 1,\dots,K-1$ is given by
    \begin{equation}
        \mathbf{x}_{(k+1)h} = \mathbf{x}_{kh}
        + h \cdot \left( \tfrac{3}{2} v_{\theta^*}(\mathbf{x}_{kh}, kh) 
        - \tfrac{1}{2} v_{\theta^*}(\mathbf{x}_{(k-1)h}, (k-1) h)  \right), 
    \label{eq:discrete_update}
    \end{equation}
    with the initialization at $k=0$ as 
    \begin{align}\label{eq:Euler_step}
		&\mathbf{x}_{h}= \mathbf{x}_{0} + h \cdot v_{\theta^{*}}(\mathbf{x}_{0},0).
	\end{align}     
    After $K$ update steps, we obtain the estimated angular-domain channel matrix $\widehat{\mathbf{H}}_{\text{ad}}^{\prime}$, i.e., the terminal state $\mathbf{x}_1$. This estimate is then converted to the complex-valued form $\widehat{\mathbf{H}}_{\text{ad}}$ via (\ref{eq:back_to_complex}) and finally mapped to the spatial domain $\widehat{\mathbf{H}}$ via (\ref{eq:back_to_sptial}). The overall inference procedure is summarized in Algorithm~\ref{alg:inference_crossmodal}.

    \subsection{Computational Complexity Analysis}
    To provide a comprehensive understanding of the practical deployment cost, we analyze the computational complexity of the proposed cross-modal flow using Big-O notation \cite{10496171}. During online channel inference, the multimodal stochastic encoder is evaluated once to generate the initial state, and the neural velocity field is invoked $K$ times along the numerical integration path. Therefore, the overall inference complexity is written as
    \begin{equation}
    	\mathcal{C}_{\mathrm{inf}}(K)=\mathcal{C}_{\mathrm{enc}}+K\mathcal{C}_{\mathrm{unet}},
    \end{equation}
	where $\mathcal{C}_{\mathrm{enc}}$ and $\mathcal{C}_{\mathrm{unet}}$ denote the complexities of the multimodal stochastic encoder and one U-Net evaluation, respectively, both of which are analyzed in the following contexts.
    
    \subsubsection{Complexity of Multimodal Stochastic Encoder}
    The total complexity of the encoder is the sum of the complexities of a CNN branch $\mathcal{C}_{\mathrm{img}}$, a PointNet branch $\mathcal{C}_{\mathrm{pc}}$, a MLP branch $\mathcal{C}_{\mathrm{gps}}$, a multi-head attention fusion module $\mathcal{C}_{\mathrm{att}}$, and two output heads $\mathcal{C}_{\mathrm{out}}$: 
    \begin{align}
    	\mathcal{C}_{\mathrm{enc}}=  \mathcal{C}_{\mathrm{img}} + \mathcal{C}_{\mathrm{pc}} + \mathcal{C}_{\mathrm{gps}} + \mathcal{C}_{\mathrm{att}} + \mathcal{C}_{\mathrm{out}}.
	\end{align}    
    For other modality combinations, the encoder complexity is obtained by removing the inactive modality-specific branches from the full model while keeping the same fusion and output.

    Specifically, the complexity of {\bf the CNN branch} is dominated by the three Conv2D layers that handle the input image of height $H$ and width $W$, and an FC layer that maps the flattened feature map to a $d$-dimensional feature vector. We let $k_{\mathrm{n}}$ represent the kernel size of the Conv2D module, and let $c_{l-1}$ and $c_l$ denote its input and output channel numbers at the $l$-th layer, respectively. Since each stride-2 downsampling operation reduces the spatial size by a factor of four, the output spatial size is $HW/4^{l}$ at the $l$-th layer. The complexity of the CNN branch is 
    \begin{equation}
    	\mathcal{C}_{\mathrm{img}} = \mathcal{O}\left(\sum_{l=1}^{3} \frac{HW}{4^l} k_{\mathrm{n}}^2 c_{l-1}c_l + \frac{HW}{64}c_3 d\right).
    \end{equation}
	{\bf For the PointNet branch}, the complexity is primarily decided by three Conv1D layers that process the input point cloud containing $U$ points, and an FC layer that maps the downsampled feature to a $d$-dimensional feature vector. We let $\kappa_\mathrm{n}$ denote the kernel size of the Conv1D module, and let $p_{l-1}$ and $p_l$ denote its input and output channel numbers at the $l$-th layer, respectively. The complexity of the PointNet branch is 
    \begin{equation}
    	\mathcal{C}_{\mathrm{pc}} = \mathcal{O}\left(U\sum_{l=1}^{3}\kappa_\mathrm{n} p_{l-1}p_l + p_3 d\right).
    \end{equation}
	{\bf For the MLP branch}, the complexity comes from two FC layers. We let $g_0$ and $g_1$ denote the dimension of sinusoidal position embedding and the hidden dimension, respectively. The complexity of the MLP branch is
    \begin{equation}
    	\mathcal{C}_{\mathrm{gps}} = \mathcal{O}\left(g_0g_1+g_1d\right).
    \end{equation}
	{\bf The multi-head attention} incurs complexity from the projections of query, key and value, the attention-score computation, and the output projection, which is expressed as
    \begin{equation}
    	\mathcal{C}_{\mathrm{att}}=\mathcal{O}(3d^2+3^2d+d^2)=\mathcal{O}(d^2).
    \end{equation}
    {\bf The two output heads} derive complexity from their Conv2D modules, which map the feature map reshaped from the $d$-dimensional fused vector to the mean and variance, respectively. As the input and output channel dimensions of both heads are 2, the total complexity of the two output heads is
    \begin{equation}
    	\mathcal{C}_{\mathrm{out}} = \mathcal{O}(2 \times N_{UE} N_{BS} \times k_{\mathrm{n}}^2 \times 2^2)=\mathcal{O}(k_{\mathrm{n}}^2 d).
    \end{equation}

	\subsubsection{Complexity of Neural Velocity Field}
	The total complexity of the time-dependent U-Net is the sum of the complexities of the encoding path $\mathcal{C}_{\mathrm{u,enc}}$, bottleneck block $\mathcal{C}_{\mathrm{u,bn}}$ and decoding path $\mathcal{C}_{\mathrm{u,dec}}$:
	\begin{align}
		\mathcal{C}_{\mathrm{unet}}=\mathcal{C}_{\mathrm{u,enc}}+\mathcal{C}_{\mathrm{u,bn}}+\mathcal{C}_{\mathrm{u,dec}}.
	\end{align}
	
	Specifically, {\bf the encoding path and the bottleneck block} incurs complexity from the two Conv2D layers and one FC layer that maps the $b$-dimensional time embedding to the input channel dimension in each block. We let $a_{l-1}$ and $a_l$ denote the input and output channel numbers of the $l$-th resolution level, respectively. Note that $a_0=2$ represents the channel number of the intermediate state $\mathbf{x}_t$. Since the first three encoder blocks progressively downsample the feature map by a factor of four, the input spatial size of the \(l\)-th block is \(d/(2\times 4^{l-1})\). Therefore, the total complexity of the encoding path and the bottleneck block is
	\begin{equation}
		\mathcal{C}_{\mathrm{u,enc}}+\mathcal{C}_{\mathrm{u,bn}} = \mathcal{O}\left( \sum_{l=1}^{4} \left[ \frac{d}{2\times 4^{l-1}} k_{\mathrm{u}}^2 \left( a_{l-1}a_l+a_l^2 \right) + ba_{l-1} \right] \right).
	\end{equation}
	{\bf For the decoding path}, the complexity is mainly attributed to the two Conv2D layers and one time-conditioning FC layer in each block. We let $a_l^{\mathrm{dec}}$ denote the output channel number of the decoder block at the $l$-th resolution level. For compact notation, we define
	\begin{equation}
		s_l=
		\begin{cases}
			a_4+a_3, & l=3,\\
			a_{l+1}^{\mathrm{dec}}+a_l, & l=1,2,
		\end{cases}
	\end{equation}
	as the input channel number at the $l$-th resolution level, which concatenates the bottleneck block output or the upsampled decoder output with the encoder skip connection. Since the Conv2D layers are applied before upsampling, the input spatial size is $d/(2\times 4^{l})$ at the $l$-th resolution level. The complexity of the decoding path is
	\begin{equation}
		\mathcal{C}_{\mathrm{u,dec}} = \mathcal{O}\left( \sum_{l=1}^{3} \left[ \frac{d}{2\times 4^{l}} k_{\mathrm{n}}^2 \left( s_la_l^{\mathrm{dec}}+(a_l^{\mathrm{dec}})^2 	\right) + bs_l \right] \right).
	\end{equation}

	\section{Experimental Results}
\begin{figure}[t]
		\centering
		\includegraphics[width = 0.8\linewidth]{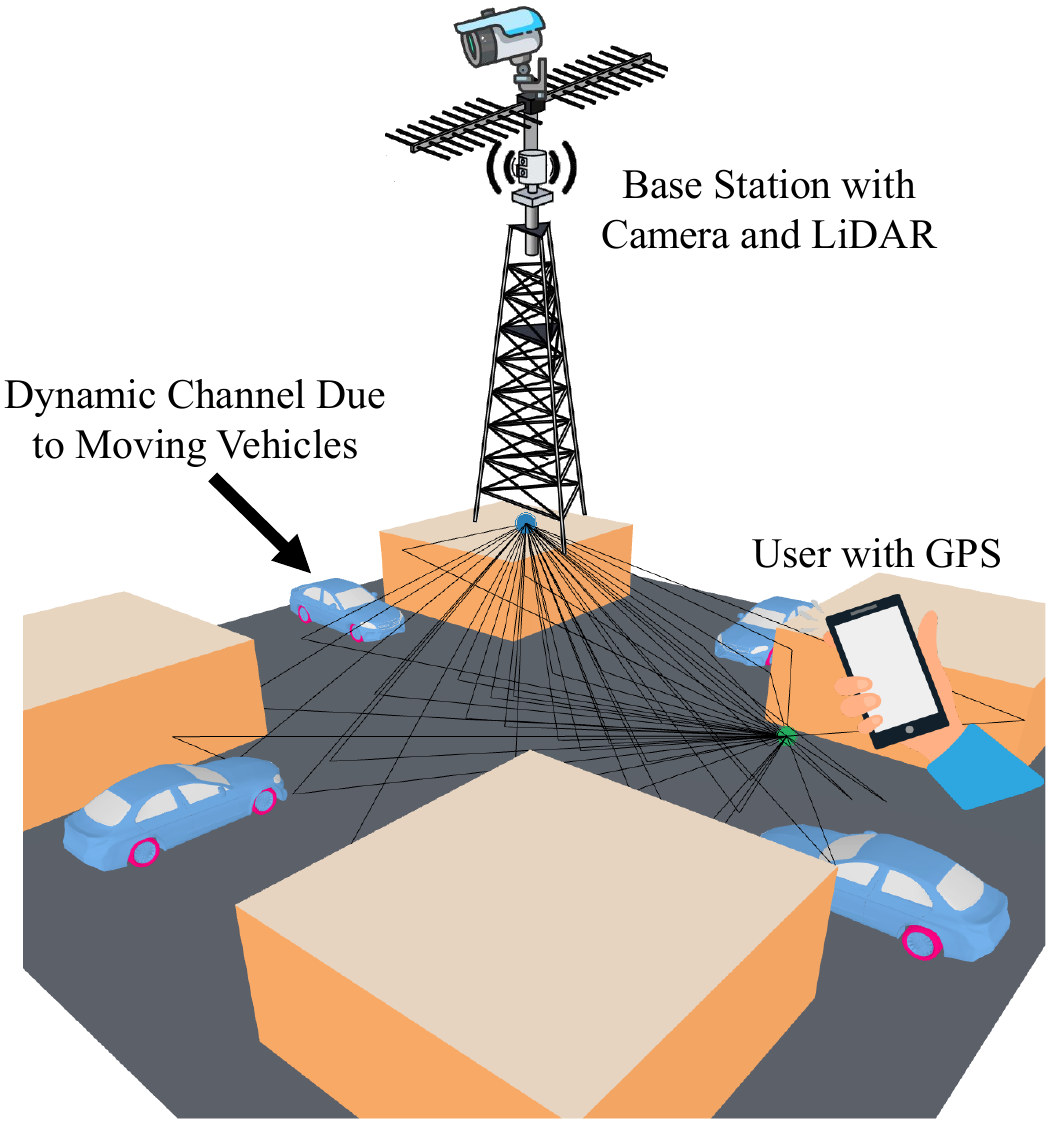}
		\caption{Illustration of the scene simulated by Sionna and Blender.}
		\label{fig:scene}
	\end{figure}

	We employ the NVIDIA Sionna link-level testbench \cite{hoydis2023} and the Blender graphics suite \cite{blender2018} as our simulators to generate 3GPP-compliant MIMO channels and multimodal sensing data. As shown in Fig. \ref{fig:scene}, we simulate an urban intersection scenario consisting of four buildings and four moving vehicles, representing the static and dynamic components of the environment, respectively. Specifically, we randomly sample $6,000$ user locations and, for each location, simulate vehicles moving across $5$ transmission frames during which the user communicates with a central BS, resulting in a dataset of $S = 30{,}000$ synchronized channel-sensing samples. Both the user and the BS employ hybrid analog and digital antenna arrays. The numbers of antennas are set to $N_{BS}$ = 64 and $N_{UE}$ = 16, and the corresponding numbers of RF chains are $N_{\text{RF}}^{BS}=4$ and $N_{\text{RF}}^{UE}=4$. Moreover, the transmission frame period is set to $T_f=10$ ms and the communication bandwidth is configured as $W $= 120 kHz, resulting in $N_{sym}=1120$ transmit symbols per frame according to the 5G standard \cite{3gpp.38.331}. 

    The proposed cross-modal flow framework is implemented in PyTorch and trained on an NVIDIA GeForce RTX 4090 24 GB GPU. The model sizes of the multimodal stochastic encoder and the neural velocity field are 209.43~MB and 142.59~MB, respectively. The dataset collected is divided into training and testing sets with a 9:1 ratio. The model is trained for $N_{\mathrm{tr}} = 2000$ epochs using a batch size of $B_s = 128$. The weight of the regularization term is $\lambda=10^{-4}$ and the learning rate of the Adam optimizer is set to $\eta=10^{-4}$.

	\subsubsection{Benchmarks}
	To evaluate the proposed framework, we compare it to representative benchmark methods. The first three benchmarks correspond to pilot-based channel estimation approaches. For these methods, the CSI acquisition period $T_{ca}$ is partitioned into two parts, with the first $(T_{ca} - T_{ce})$ used for pilot transmission and the remaining $T_{ce}$ used for executing channel estimation algorithms. This implies that the number of pilot symbols is given by $N_p=(T_{ca} - T_{ce}) / T_f \times N_{\text{sym}}$. Since each transmitted pilot symbol provides $N_{\text{RF}}^{BS}$ independent measurements, the pilot density for channel estimation is calculated as $\alpha= (N_p \times N_{\text{RF}}^{BS})/(N_{UE}\times N_{BS})$.

	\noindent $\bullet$ \textbf{LS-Based Channel Estimation \cite{kay1993fundamentals}:}
	The least-squares (LS) method is employed to estimate the channel coefficient matrix directly from the received pilot signals. The computational complexity of LS is dominated by the singular value decomposition-based pseudo-inverse calculation and the subsequent matrix-vector multiplication.
	
	\noindent $\bullet$ \textbf{LASSO-Based Channel Estimation \cite{7094443}:}
	The LASSO estimator is applied to estimate the angular-domain channel matrix from the received pilots, which is then transformed back into the spatial domain. The computational complexity of LASSO is primarily determined by iterative sparse recovery over the angular-domain channel dictionary, where each iteration involves forward and adjoint multiplications with the effective pilot-dictionary sensing matrix, followed by shrinkage updates.
	
	\noindent $\bullet$ \textbf{DDPM-Based Channel Estimation \cite{10930691}:} 
	A diffusion denoiser implemented by a lightweight time-dependent CNN is first pretrained on angular-domain channel samples and then used as a generative prior for channel estimation. During inference, the reverse diffusion process starts from Gaussian noise and progressively refines the sample within the denoising diffusion probabilistic model (DDPM) framework. This refinement is guided by posterior scores derived from both the pretrained channel prior and the received pilots. After the reverse diffusion iterations, the estimated angular-domain channel is transformed back to the spatial domain. The computational complexity mainly comes from two parts, namely the posterior-score calculation and the repeated evaluations of the diffusion denoiser across reverse diffusion iterations.

	\noindent $\bullet$ \textbf{DDIM-Based Channel Estimation \cite{10615282}:} 
	This scheme follows the same pipeline as the DDPM-based method above, but adopts the so-called denoising diffusion implicit model (DDIM) sampling in place of DDPM sampling. Specifically, DDIM employs a non-Markovian reverse process that allows the sampler to skip redundant intermediate steps and follow a more efficient denoising trajectory, thereby achieving higher channel estimation accuracy under the same sampling-step budget. As in the DDPM-based method, the computational complexity derives from posterior-score calculation and repeated denoiser evaluations.
	
	We further consider a benchmark that realizes environment-aware channel inference based on the user's location. 
	
	\noindent $\bullet$ \textbf{KNN-Based Channel Inference \cite{9473871}:} The channels generated by Sionna are represented using a ray-tracing model characterized by the path gain, phase, and zenith/azimuth angles of departure (AoDs) and arrival (AoAs). For each user location, the three strongest propagation paths are first selected from each transmission frame. The corresponding channel parameters are then averaged across five consecutive frames to obtain three representative paths. Based on the training samples, an inverse-distance-weighted KNN algorithm is employed to predict the path parameters at unseen locations according to their spatial proximity to the training locations, which are subsequently used for reconstructing the channel coefficient matrix. The computational complexity mainly comes from the nearest-neighbor search over the training user locations and the subsequent channel reconstruction using the predicted dominant path parameters.
	
	The aforementioned benchmarks are also extended to downstream physical-layer operations that rely on CSI, among which we specifically consider beamforming design in our experiments and include the following task-oriented benchmark.
	
	\noindent $\bullet$ \textbf{Multimodal-Sensing-Aided Beam Selection \cite{10539181}:}
	The method in~\cite{10539181} considers a multiple-input single-output (MISO) system using a fully digital antenna array. To adapt it to our MIMO setting with hybrid analog-digital arrays, we construct DFT codebooks for the analog transmit precoder (TPC) and receive combiner (RC), and Grassmannian codebooks for their digital counterparts. A Transformer network is then employed to infer the optimal beam indices in these codebooks from multimodal sensing inputs (i.e., RGB image, LiDAR point cloud, and user coordinate). The final hybrid TPC and RC are obtained by combining the selected analog and digital components. The computational complexity mainly comes from multimodal feature extraction and Transformer-based prediction, where the former follows a similar complexity composition to our multimodal stochastic encoder in Sec.~IV-D and the latter is determined by self-attention and feed-forward operations in each Transformer layer.

	\begin{figure}[t]
		\centering 
		\subfigure[Impact of $K$ on NMSE]{\includegraphics[width=0.241\textwidth]{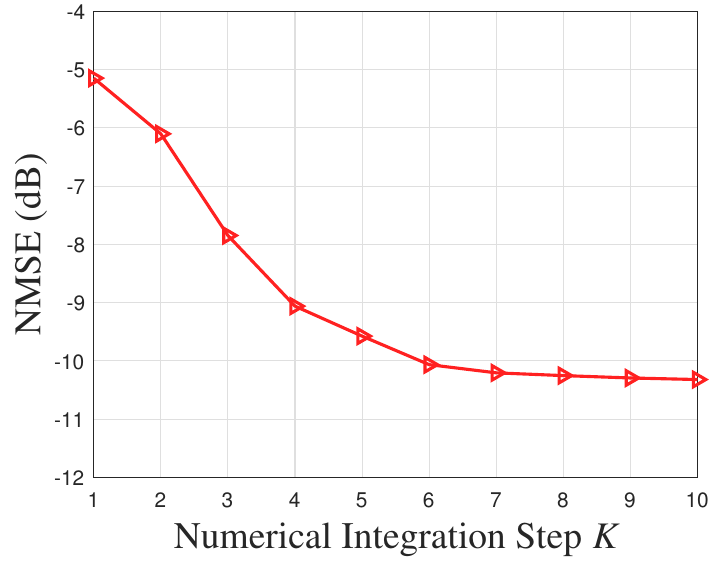}}
		\subfigure[Impact of $K$ on cosine similarity]{\includegraphics[width=0.241\textwidth]{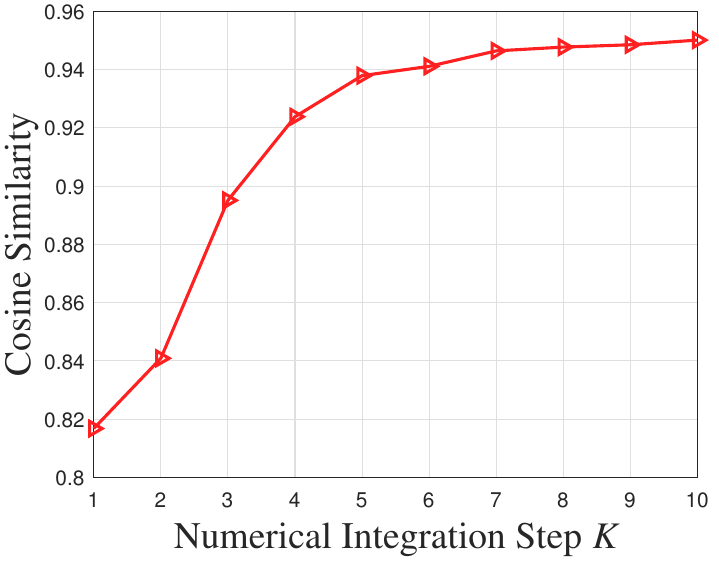}} 
		\caption{The impact of numerical integration step $K$ on channel estimation accuracy.}
		\label{fig:Accuracy_over_K}
		\vspace{0.2in}
	\end{figure} 

    \subsubsection{Evaluation Metrics} 
    Given the estimated channel matrix $\widehat{\mathbf{H}}^{[i]}$ and its ground-truth $\mathbf{H}^{[i]}$ at the $i$-th transmission frame, the channel estimation accuracy is evaluated using the normalized mean square error (NMSE), defined as
	\begin{equation}
				\textbf{(NMSE)} \quad  \frac{ \left\| \mathbf{H}^{[i]} - \widehat{\mathbf{H}}^{[i]} \right\|_{F}^2}
				{\left\| \mathbf{H}^{[i]} \right\|^2_{F}},
	\end{equation} 
	While NMSE quantifies the element-wise estimation error, we also consider the cosine similarity for richer characterization. This metric quantifies the alignment between the subspaces spanned by the estimated and ground-truth channel matrices, which is particularly relevant for beamforming-oriented tasks: 
	\begin{equation}
				\textbf{(Cosine Similarity)}  \quad
				\frac{ \|(\mathbf{H}^{[i]})^{H} \widehat{\mathbf{H}}^{[i]}\|_F}{\|\mathbf{H}^{[i]}\|_F \|\widehat{\mathbf{H}}^{[i]}\|_F}.
	\end{equation}	 
	Both metrics are averaged over all transmission frames and user locations.

	To evaluate the beamforming performance, we consider the achievable spectral efficiency (SE),  which depends on the channel acquisition time duration \(T_{ca}\). Explicitly, \(T_{ca}\) is defined as the duration from the beginning of channel acquisition until the beam selection is completed. Although this definition slightly extends the conventional notion of channel acquisition, it captures the total latency in inferring the beamforming vectors. For fair comparison, we use the same analog and digital codebooks as those in the multimodal-sensing-aided beam selection benchmark, and adopt the two-stage search algorithm \cite{8007256} to obtain the hybrid TPC $\widehat{\mathbf{F}}^{[i]}\in\mathbb{C}^{N_{BS}\times 1}$ and RC $\widehat{\mathbf{W}}^{[i]}\in\mathbb{C}^{N_{UE}\times 1}$ that align with the dominant spatial direction of the channel. Accordingly, the instantaneous SE is 
	\begin{align}
	    R(\widehat{\mathbf{W}}^{[i]},\widehat{\mathbf{F}}^{[i]},\mathbf{H}^{[i]}) = \log_2  \Big(1+\gamma_{SNR}\left|(\widehat{\mathbf{W}}^{[i]})^{H} \mathbf{H}^{[i]}\widehat{\mathbf{F}}^{[i]} \right|^2\Big),
	\end{align}          
	where $\gamma_{SNR}$ is the signal-to-noise ratio (SNR). 
	
	In pilot-based channel estimation schemes, both pilot signaling and CSI recovery are performed within the channel acquisition period $T_{ca}$, during which no data transmission occurs.
	The resultant effective achievable SE within one frame is given by 
	     	\begin{align}\label{eq:average_rate_ce}
	             \textbf{ (Pilot-Based SE)} \quad \frac{T_{f}-T_{ca}}{T_f}R(\widehat{\mathbf{W}}^{[i]},\widehat{\mathbf{F}}^{[i]},\mathbf{H}^{[i]}).
	   		\end{align} 
	
	By contrast, in environment-aware and sensing-aided communications, as illustrated in Fig.~\ref{fig:Frame_Stucture}, the computation processes for the current frame are performed concurrently with data transmission using the beamformers from the previous frame. 
	The average achievable SE within a frame is therefore calculated as
	\begin{align}\label{eq:average_rate_ci}
		&  \textbf{ (Sensing-Aided SE)} \quad \frac{T_{ca}}{T_f}R(\widehat{\mathbf{W}}^{[i-1]},\widehat{\mathbf{F}}^{[i-1]},\mathbf{H}^{[i]})-\frac{128\text{ }\text{bits}}{T_f W}\notag\\
		&\qquad\qquad\qquad\qquad\qquad~~ +\frac{T_f-T_{ca}}{T_f} R(\widehat{\mathbf{W}}^{[i]},\widehat{\mathbf{F}}^{[i]},\mathbf{H}^{[i]}),
	\end{align} 
	where the second term represents the overhead associated with GPS query, assuming that 128 bits are used to encode the two-dimensional user coordinate. All metrics are averaged over the last four transmission frames, excluding the first frame to avoid initialization ambiguity, and over all user locations.

	\begin{figure}[t]
		\centering
		\includegraphics[width = 1\linewidth]{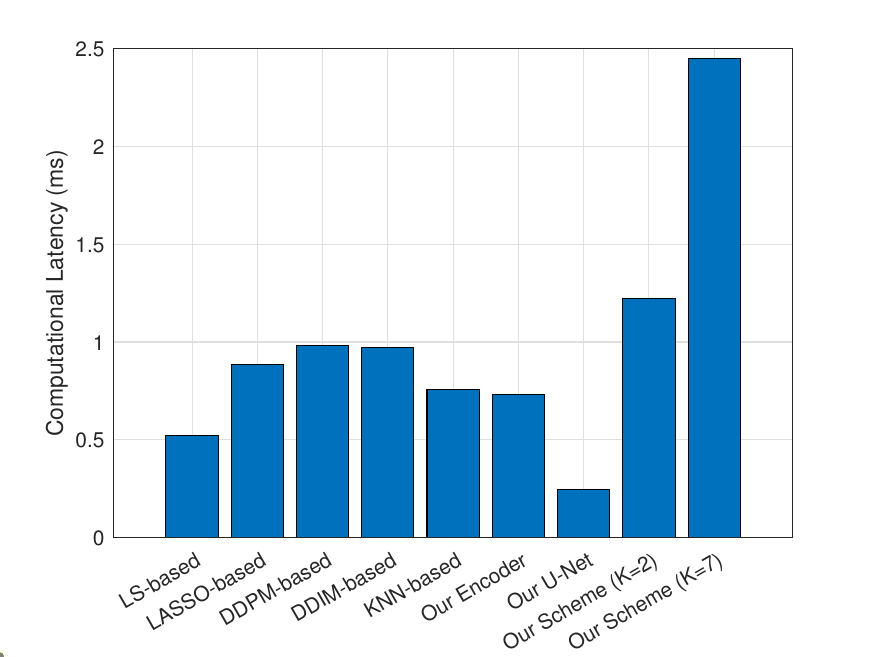}
		\caption{Computational latency comparison of sensing-aided channel inference and pilot-based channel estimation schemes.}
		\label{fig:Latency}
	\end{figure}

    \subsection{Parameter and Ablation Studies}
    We study the impact of the number of numerical integration steps $K$ used during inference regardless of the latency constraint, which determines the discretization resolution of the flow matching process. As shown in Fig.~\ref{fig:Accuracy_over_K}, the channel estimation performance improves as $K$ increases, with the NMSE decreasing and the cosine similarity increasing due to the reduced numerical integration error. However, when $K$ exceeds $7$, further increasing it to $10$ yields only a marginal NMSE improvement of $0.5$~dB and approximately $0.01$ in cosine similarity gain. This indicates that the discrete update already provides a sufficiently accurate approximation of the continuous integration process. Based on this observation, we set $K=7$ as the default value in the following experiments, unless otherwise specified.
    
    The number of numerical integration steps $K$ directly determines the inference latency of the proposed cross-modal flow framework. Accordingly, Fig.~\ref{fig:Latency} reports the computational latency of all the channel estimation and inference schemes considered. The multimodal stochastic encoder is activated once per inference, requiring approximately $0.75$~ms, while the U-Net is harnessed $K$ times, with an average runtime of $0.25$~ms per step. Consequently, the total inference latency of our method is about $1.25$ ms for $K=2$ and remains below $2.5$ ms for $K=7$, both of which are comparable to the typical CSI acquisition period. For the sensing-aided channel inference schemes, the CSI acquisition period is equal to the sum of the GPS query duration and the channel inference latency. Given that the GPS query time is negligible, the overall latency of our proposed framework remains on the same order of magnitude as the latency budget of typical communication systems. Moreover, for the pilot-based channel estimation schemes (i.e., LS, LASSO, DDPM and DDIM), the CSI acquisition period corresponds to the overall end-to-end acquisition budget including the pilot transmission stage and the execution stage of the channel estimation algorithms. Unlike the single-step LS and LASSO estimators, which achieve sub-millisecond computation latency, DDPM and DDIM are iterative methods whose computation time scales with the number of sampling iterations. To ensure a fair comparison under the same real-time latency budget and to reserve sufficient time for pilot transmission within each CSI acquisition period, we set the channel estimation period to $T_{ce}=1$~ms in the following experiments, while the corresponding pilot transmission period is $T_{ca}-1$~ms.
 
	\begin{figure}[t]
	 	\centering
	 	\subfigure[Attention Matrix Visualization]{\includegraphics[width=0.255\textwidth]{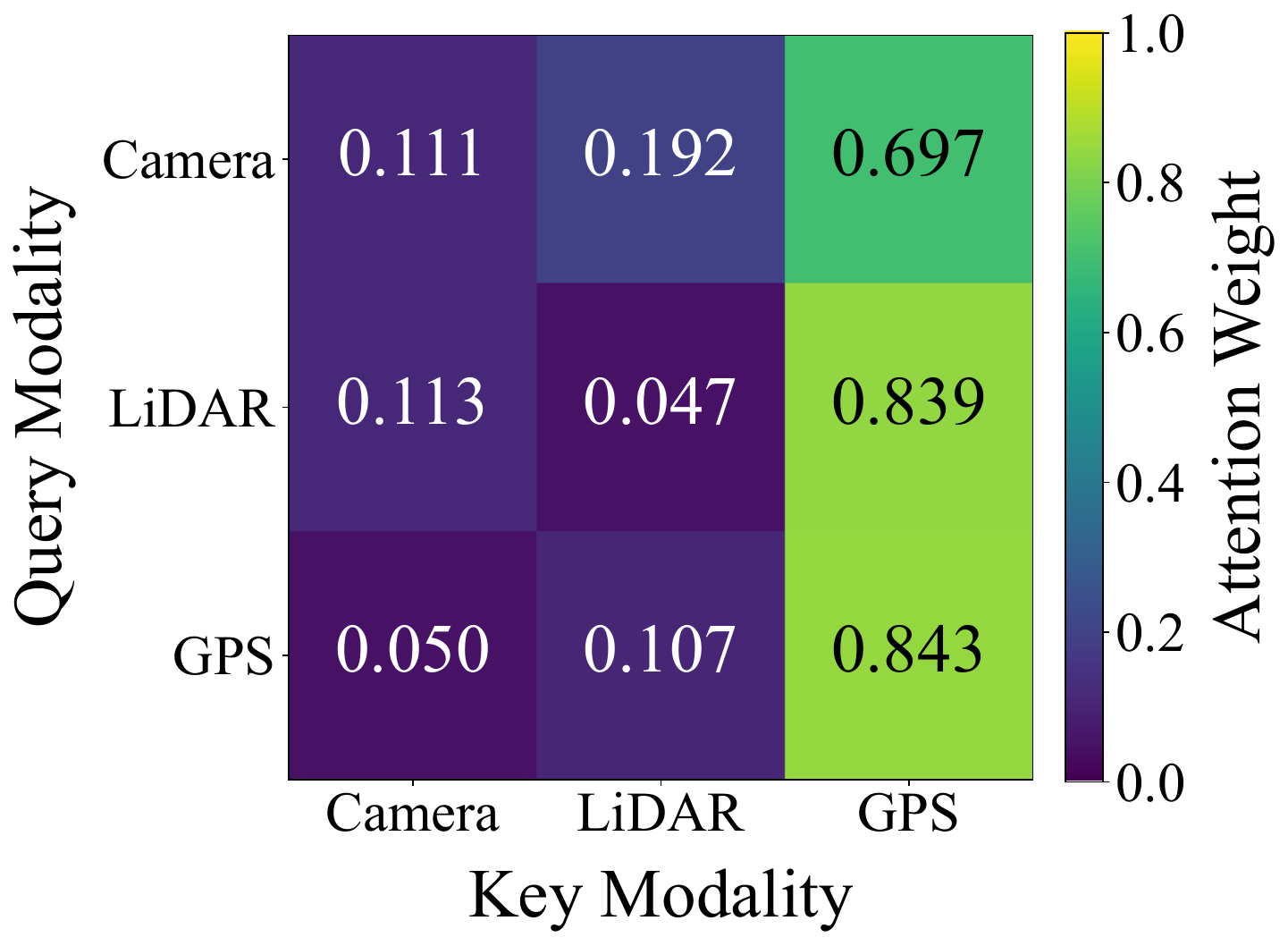}}
	 	\subfigure[Feature-Occlusion Dependency]{\includegraphics[width=0.225\textwidth]{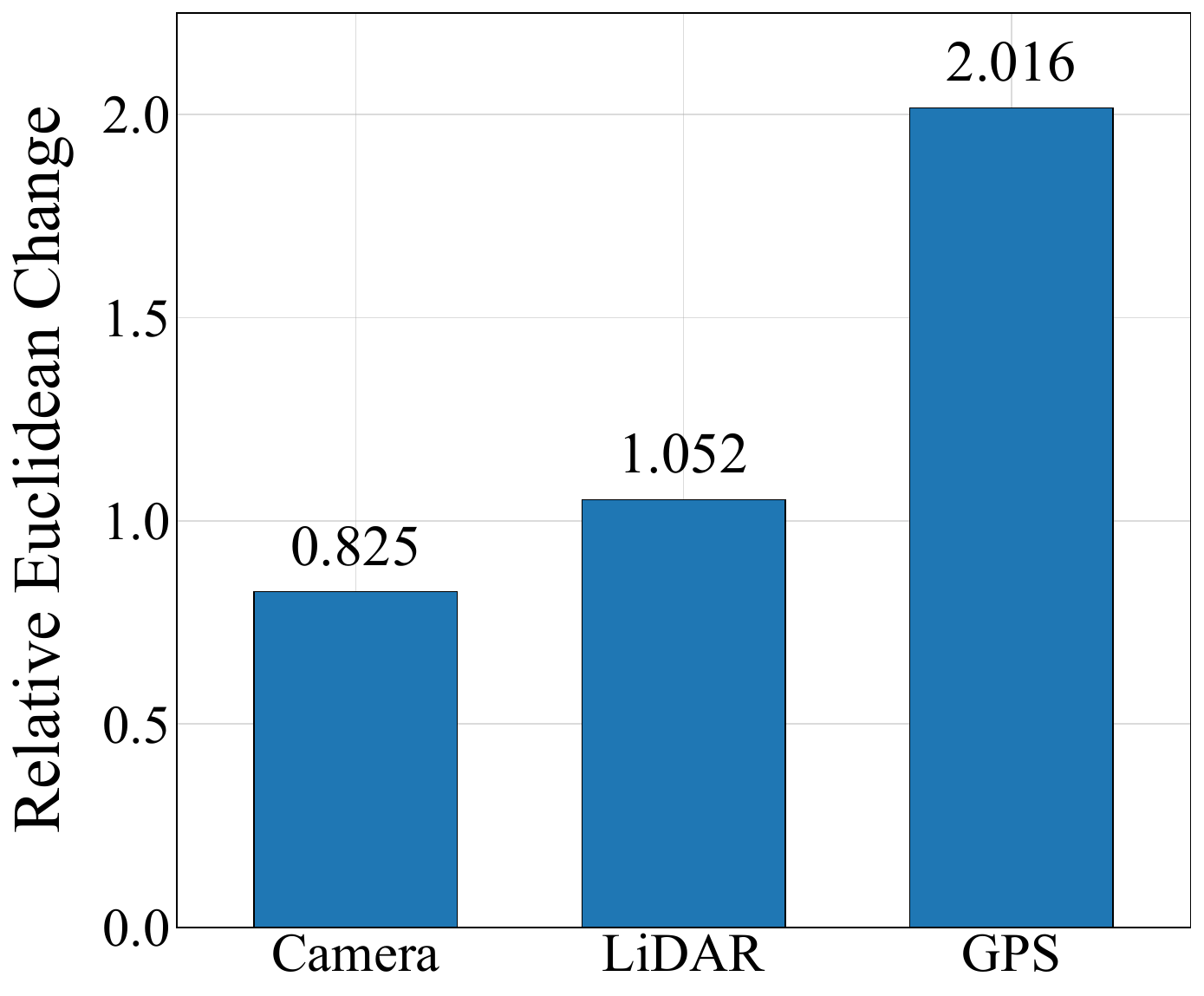}} 
	 	\caption{The dependency of fused representation on three individual features.}
	 	\label{fig:test_dependency}
	 	\vspace{0.2in}
	 \end{figure}  
 
    To improve the interpretability of the proposed multimodal fusion mechanism, Fig.~\ref{fig:test_dependency} examines the dependency of the fused representation on the three individual modality features. Specifically, Fig.~\ref{fig:test_dependency}(a) visualizes the attention matrix averaged over all test samples and attention heads. The results show that GPS receives the highest attention weights of 0.697, 0.843, and 0.839 from the Camera, GPS, and LiDAR queries, respectively, which indicates that the GPS representation serves as the primary information reference during the fusion process. Meanwhile, the non-zero weights assigned to Camera and LiDAR suggest that they also provide complementary information. Moreover, Fig.~\ref{fig:test_dependency}(b) presents a feature-occlusion analysis, where we individually set the Camera, GPS, or LiDAR feature vector to zero while keeping the other two modalities unchanged, and then measure the relative Euclidean change in the fused representation. The results demonstrate that removing the GPS feature produces the largest Euclidean change of 2.016, compared with 1.052 for LiDAR and 0.825 for Camera. This further confirms that the fused representation is most sensitive to GPS, while the LiDAR and Camera solutions provide additional complementary information. 
    	
	\begin{table}[t]
		\centering
		\caption{Ablation Studies}
		\small
		\setlength\tabcolsep{5pt}
		\begin{tabular}{lcc}
			\toprule
			Ablation Variant  & NMSE~(dB)~$\downarrow$  & Cosine Similarity~$\uparrow$ \\
			\midrule
			\multicolumn{3}{c}{\textit{Ablation on Loss Objective}} \\
			\midrule
			CFM        & -4.86  & 0.82 \\
			CFM+Recon.  & -5.78  & 0.86 \\
			CFM+MA     & \textbf{-10.21} & \textbf{0.95} \\
			\midrule
			\multicolumn{3}{c}{\textit{Ablation on Inference Module}} \\
			\midrule
			Enc. Only w/ CFM+MA  & 8.89 & 0.80 \\
			Enc. Only w/ MSE  &  -6.98 & 0.89 \\
			Enc. + Lightweight Reg.  & -8.32 & 0.92 \\
			Enc. + ODE Integration & \textbf{-10.21} & \textbf{0.95} \\
			\midrule
			\multicolumn{3}{c}{\textit{Ablation on Sensing Modality}} \\
			\midrule
			GPS Only  & -3.82 & 0.78 \\
			Camera Only & 0.30 & 0.01 \\
			LiDAR Only & 0.24 & 0.02 \\
			Camera+LiDAR & 0.17 & 0.05 \\
			GPS+Camera & -9.13 & 0.93 \\
			GPS+LiDAR & -9.38 & 0.93 \\
			GPS+Camera+LiDAR & \textbf{-10.21} & \textbf{0.95} \\
			\bottomrule
		\end{tabular}
		\label{tab:Ablation_Loss_Inference}
	\end{table}
    
    In Table~\ref{tab:Ablation_Loss_Inference}, we conduct ablation studies to comprehensively evaluate the impact of the training objective, inference module, and sensing modality. First, to justify the rationale for including the modality alignment loss, we compare three training objectives: the CFM baseline trained solely with the flow matching loss; the CFM+Recon. variant that introduces modality-specific decoders to reconstruct each modality from a shared latent representation following \cite{liu2024flowing}, and thus incorporates the reconstruction loss and the KL regularization; and the proposed CFM+MA scheme that jointly optimizes the CFM and modality alignment losses. It is observed that the CFM+MA scheme achieves the best overall performance, yielding an NMSE of -$10.21$~dB and a cosine similarity of $0.95$. The CFM baseline exhibits the worst estimation quality, with an NMSE of -$4.86$~dB and a cosine similarity of $0.82$, while the CFM+Recon. configuration moderately improves the performance to -$5.78$~dB NMSE and $0.86$ cosine similarity. These results suggest that, compared with merely preserving multimodal information in the latent space as in the CFM+Recon. scheme, enforcing explicit alignment between the source and target distributions improves transport efficiency and thus enhances channel estimation accuracy.
    
    To further justify the necessity of ODE integration, we compare four configurations of the inference module: the Enc. Only w/ CFM+MA baseline, which directly treats the multimodal encoder output as the channel prediction after training the encoder with the proposed CFM+MA objective; the Enc. Only w/ MSE scheme, which also directly uses the encoder output as the final channel estimate, but trains the encoder with the MSE loss; the Enc. + Lightweight Reg. variant, which appends a lightweight CNN regressor to the encoder and jointly optimizes them using MSE loss; and our proposed Enc. + ODE Integration scheme. As shown in the Table~II, solely using the encoder trained with CFM+MA leads to the worst NMSE of $8.89$~dB but still achieves a relatively high cosine similarity of $0.80$, because the encoder output mainly serves as a direction-aligned latent representation rather than a calibrated channel estimate. After replacing CFM+MA with MSE supervision, the NMSE improves significantly to -6.98~dB and the cosine similarity increases to 0.89, which suggests that direct regression makes the encoder output similar to the channel space. Further introducing the lightweight CNN improves the NMSE and cosine similarity to $-8.32$~dB and $0.92$, respectively, but still remains inferior to the proposed Enc. + ODE Integration scheme, which achieves the best NMSE of $-10.21$~dB and cosine similarity of $0.95$. This is because the lightweight CNN learns a shortcut from the latent space to the channel space, which may be insufficient for capturing the complex distributional discrepancy.

    \begin{figure}[t]
    	\centering
    	\begin{minipage}[b]{0.165\textwidth}
    		\centering
    		\includegraphics[width=\linewidth]{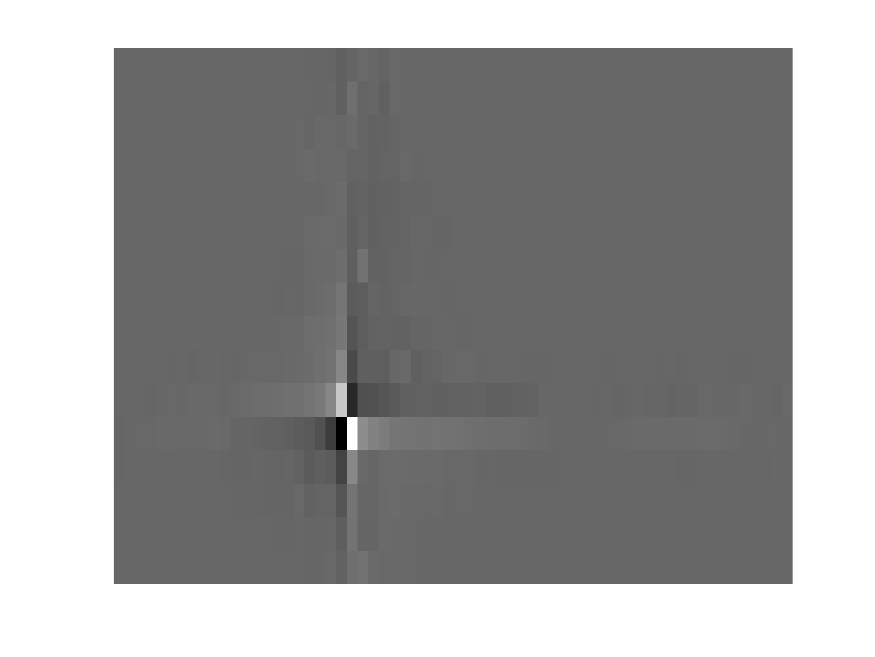}
    		{\footnotesize (a)  k=0, CosSim=0.90,\\ ~~~NMSE= 2.88 dB}
    	\end{minipage}\hspace{-0.2em}%
    	\begin{minipage}[b]{0.165\textwidth}
    		\centering
    		\includegraphics[width=\linewidth]{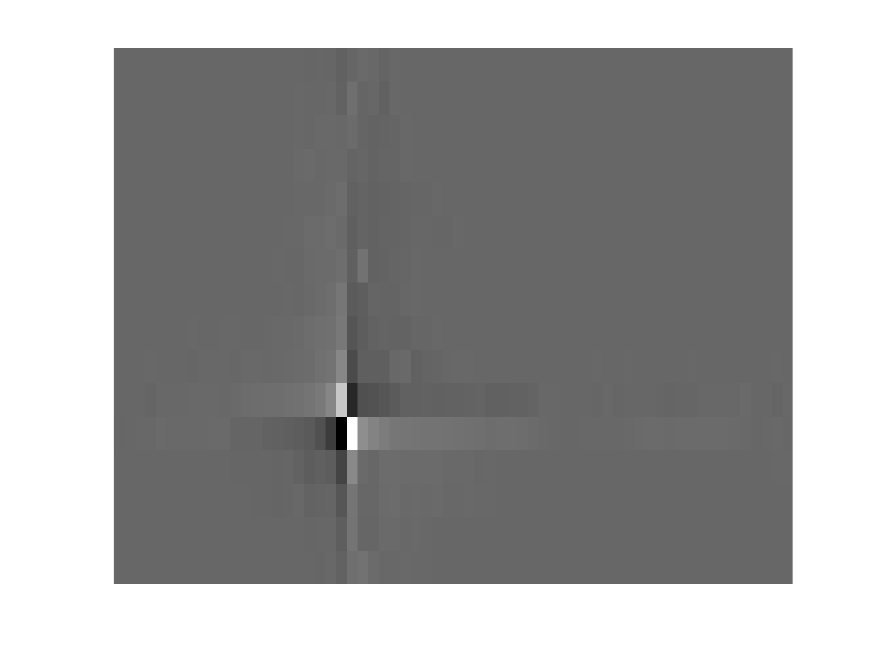}
    		{\footnotesize  (b) k=1, CosSim=0.91, \\ ~~NMSE= 1.48 dB}
    	\end{minipage}\hspace{-0.2em}%
    	\begin{minipage}[b]{0.165\textwidth}
    		\centering
    		\includegraphics[width=\linewidth]{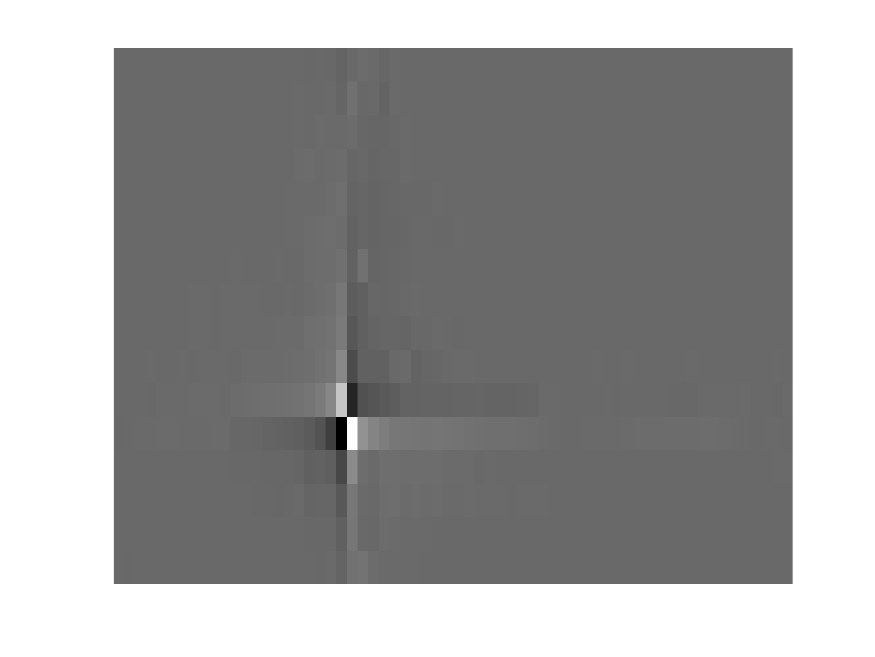}
    		{\footnotesize (c) k=3, CosSim=0.93, \\ ~~NMSE= -2.19 dB}
    	\end{minipage}
    	\vspace{0.05em} 
    	\begin{minipage}[b]{0.165\textwidth}
    		\centering
    		\includegraphics[width=\linewidth]{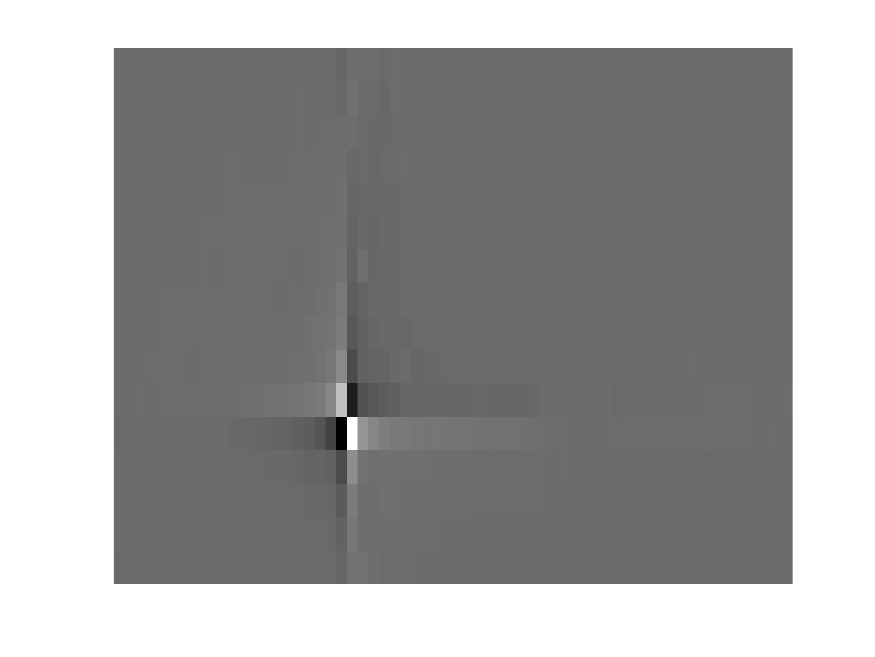}
    		{\footnotesize  (d) k=5, CosSim=0.95, \\ ~~~NMSE=-8.04 dB}
    	\end{minipage}\hspace{-0.2em}%
    	\begin{minipage}[b]{0.165\textwidth}
    		\centering
    		\includegraphics[width=\linewidth]{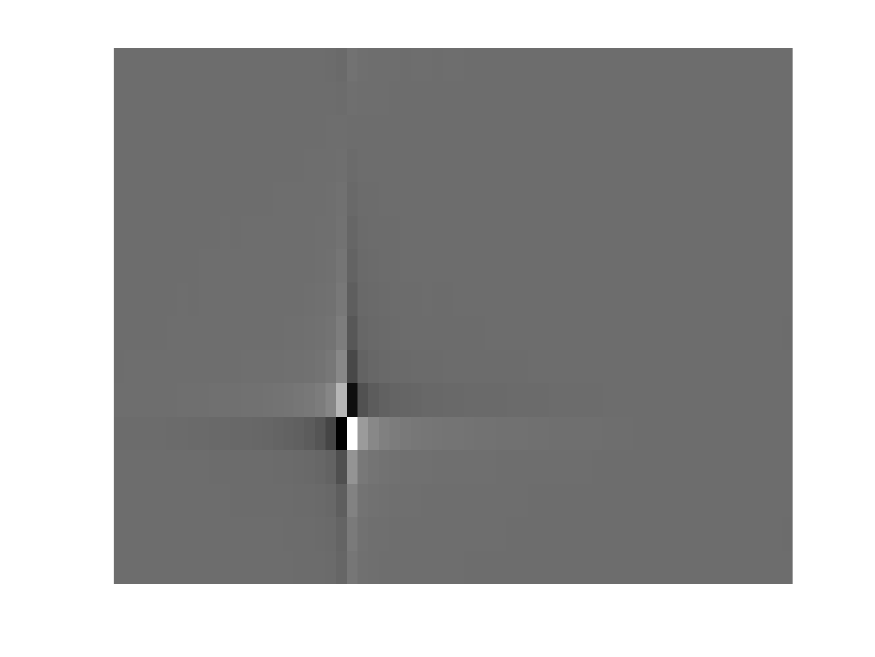}
    		{\footnotesize  (e) k=7, CosSim=0.96, \\ ~~~NMSE=-10.49 dB}
    	\end{minipage}\hspace{-0.2em}%
    	\begin{minipage}[b]{0.165\textwidth}
    		\centering
    		\includegraphics[width=\linewidth]{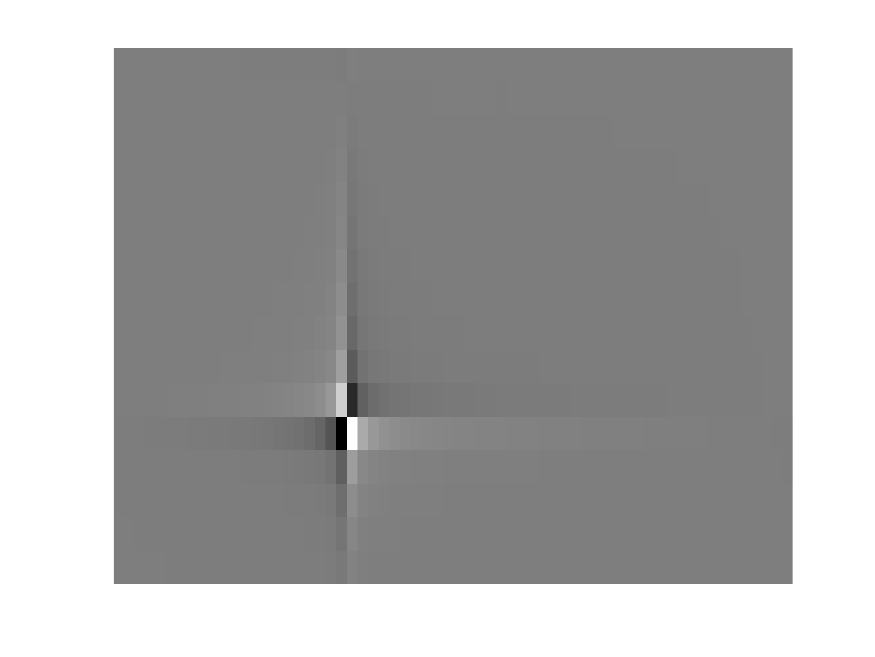}
    		{\footnotesize (f) Ground-Truth Angular \\ Domain Channel Matrix}
    	\end{minipage}
    	\caption{\footnotesize Cross-modality refinement process from the aligned multimodal sensing latent representation to the angular-domain channel matrix. }
    	\label{fig:evolution_process}
    \end{figure}

    \begin{figure}[t]
    	\centering
    	\subfigure[Impact of SNR on NMSE]{\includegraphics[width=0.45\textwidth]{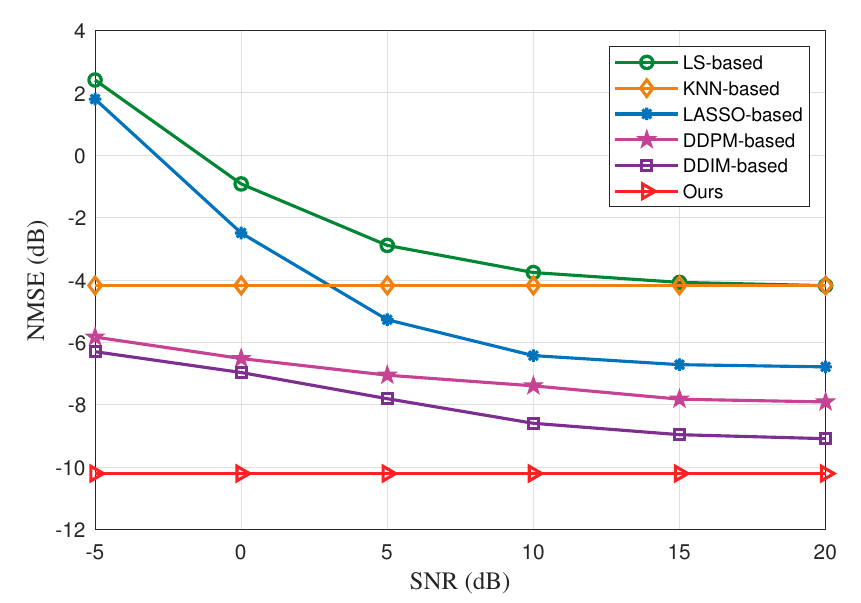}}
    	\subfigure[Impact of SNR on cosine similarity]{\includegraphics[width=0.45\textwidth]{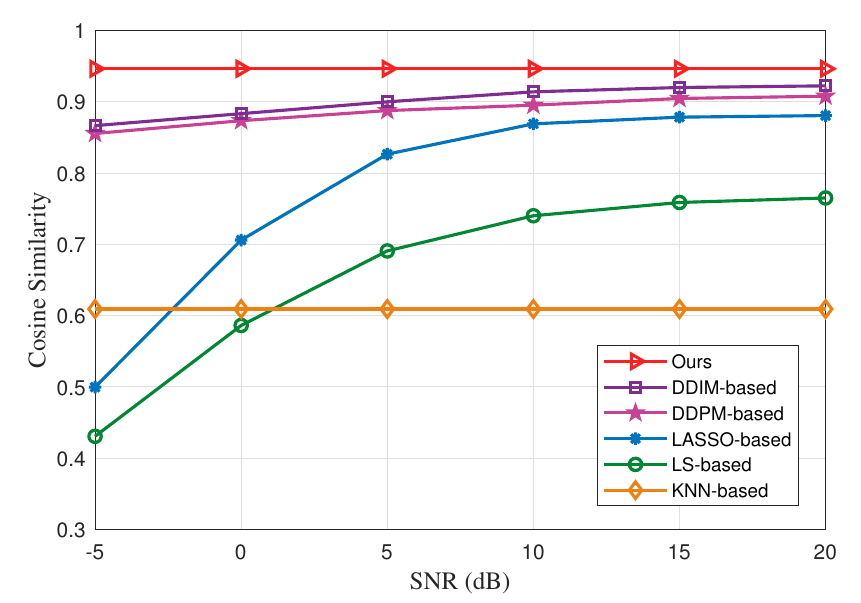}} 
    	\caption{The impact of SNR on channel estimation accuracy under the CSI acquisition period $T_{ca}$ = 2.5 ms.}
    	\label{fig:Accuracy_over_SNR}
    	\vspace{0.2in}
    \end{figure}
    
    \begin{figure}[t]
    	\centering
    	\subfigure[Impact of $T_{ca}$ on NMSE]{\includegraphics[width=0.45\textwidth]{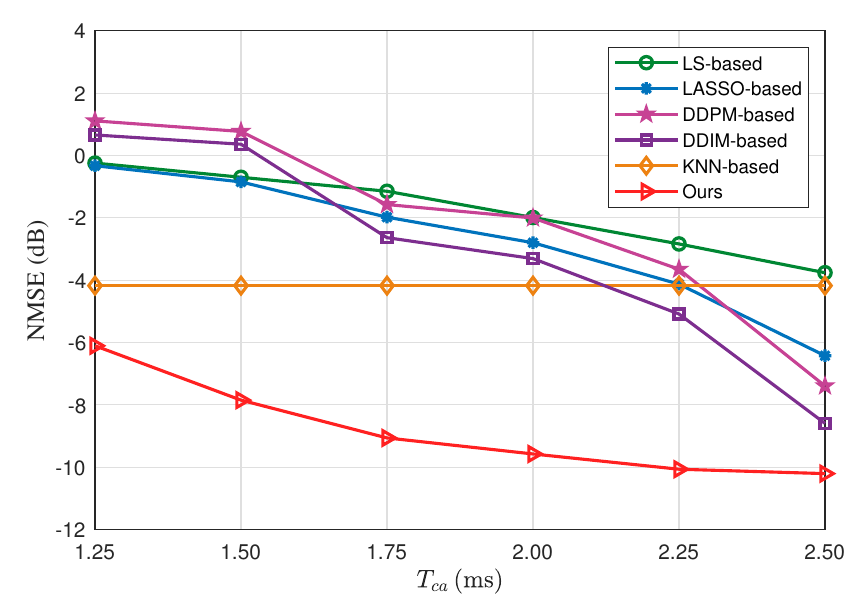}}
    	\subfigure[Impact of $T_{ca}$ on cosine similarity]{\includegraphics[width=0.45\textwidth]{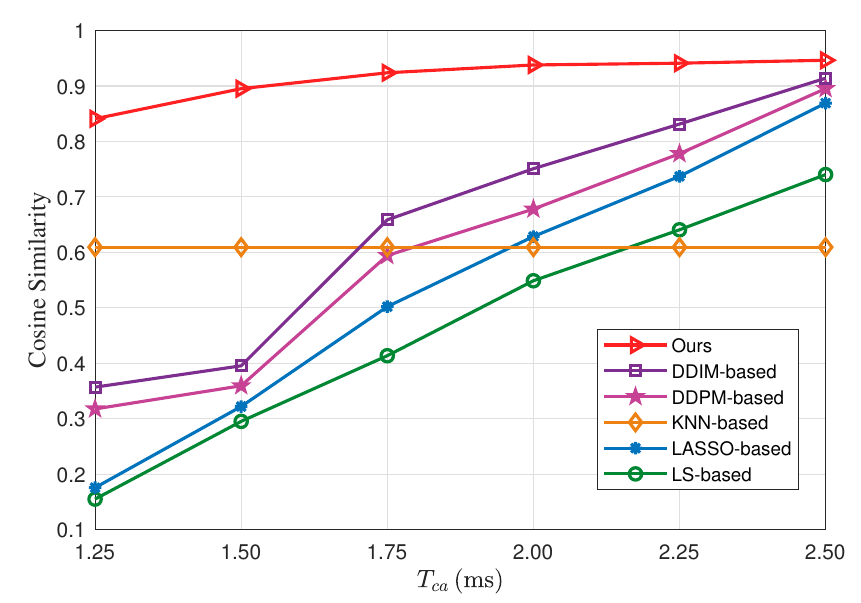}} 
    	\caption{The impact of CSI acquisition period $T_{ca}$ on channel estimation accuracy under the SNR = 10 dB. The numerical integration step $K$ varies with $T_{ca}$.}
    	\label{fig:Accuracy_over_Ratio}
    	\vspace{0.2in}
    \end{figure}
    
    Finally, we examine the effect of different sensing modalities. Among the single-modality settings, GPS-only achieves substantially better performance than camera-only and LiDAR-only, with an NMSE of $-3.82$~dB and a cosine similarity of 0.78. This advantage arises because GPS directly encodes the user-BS geometry and provides a location-dependent prior that implicitly captures the static propagation context. In contrast, camera and LiDAR mainly characterize the surrounding environment and lack explicit user-BS geometric information when used independently, making them less informative for channel inference than GPS-only. When GPS is combined with either camera or LiDAR, the performance improves markedly, with GPS+Camera achieving an NMSE of $-9.13$~dB and a cosine similarity of 0.93, and GPS+LiDAR achieving an NMSE of $-9.38$~dB and the same cosine similarity of 0.93. This demonstrates that both modalities can complement GPS by providing dynamic environmental information. By contrast, Camera+LiDAR performs poorly, with an NMSE of 0.17~dB and a cosine similarity of 0.05, because neither of them provides the essential user-BS geometric relationship required to characterize wireless propagation. Furthermore, incorporating a Camera into the GPS+LiDAR configuration yields the best estimation performance, improving the NMSE from -9.38~dB to -10.21~dB and the cosine similarity from 0.93 to 0.95. Although GPS and LiDAR already provide a geometrically near-complete representation of the propagation environment, the camera supplies additional visual and semantic information that is not fully captured by LiDAR point clouds, such as object categories, surface appearance, and other visual cues related to the propagation environment. This complementary information enables a more comprehensive characterization of the dynamic environment, thereby improving CSI inference accuracy. 
    
    Overall, the results above reveal a practical cost-performance tradeoff. The lower-cost GPS+Camera configuration may be attractive for cost-sensitive deployments, whereas the full GPS+Camera+LiDAR configuration is more suitable for critical scenarios such as high-capacity hotspots, smart intersections, and roadside units, where the improved channel inference accuracy can justify the additional LiDAR cost.    

    \subsection{Evaluation on Channel Estimation Accuracy}
    Fig.~\ref{fig:evolution_process} visualizes the cross-modality refinement process from the aligned multimodal sensing latent representation to the angular-domain channel matrix during inference. At the initial step (i.e., $k=0$), the latent representation has already been aligned with the channel domain and captures the dominant angular support pattern of the ground-truth channel, achieving an NMSE of $2.88$~dB and a cosine similarity of $0.90$. Nevertheless, the initial state still contains noticeable diffuse components and streak-like artifacts around the dominant angular paths. As the neural ODE is numerically integrated forward, i.e., as $k$ increases, the estimated channel undergoes a progressive refinement process, during which spurious components are gradually suppressed and energy becomes increasingly concentrated around the dominant angular paths. Correspondingly, the NMSE steadily decreases while the cosine similarity increases. Upon completing the integration at $k=7$, the terminal state exhibits the cleanest and most distinct sparsity pattern, closely resembling the ground-truth angular-domain channel matrix in Fig.~\ref{fig:evolution_process}(f), with an NMSE of $-10.49$~dB and a cosine similarity of $0.96$. This evolution highlights the different yet coordinated roles of the two components, whereby the multimodal encoder provides a structured and channel-aligned starting point, while the flow model progressively purifies and calibrates it toward the target channel realization.
	
	\begin{figure}[t]
		\centering
		\includegraphics[width = 0.93\linewidth]{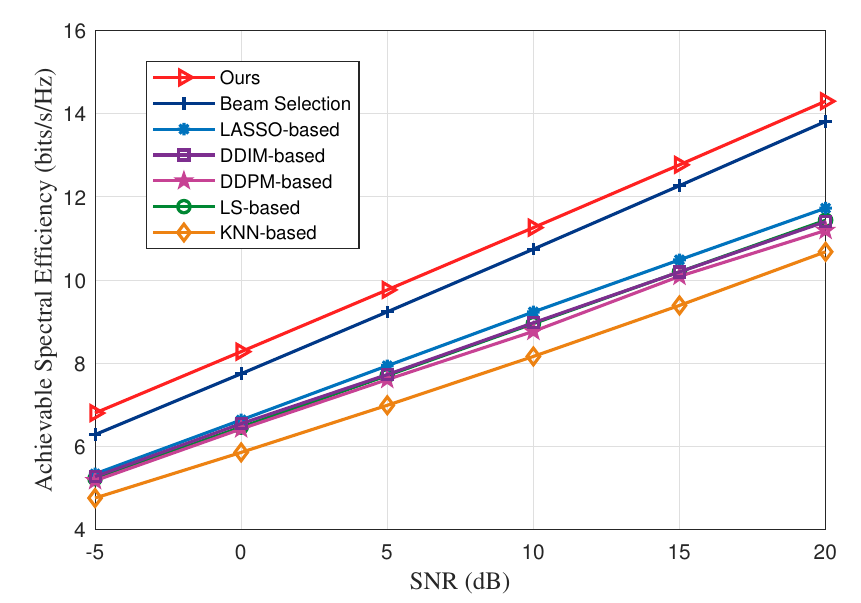}
		\caption{The impact of SNR on achievable SE under the CSI acquisition period $T_{ca}$ = 2.5 ms.}
		\label{fig:Rate_over_SNR}
	\end{figure}

    In Fig.~\ref{fig:Accuracy_over_SNR}(a) and Fig.~\ref{fig:Accuracy_over_SNR}(b), we investigate the effect of SNR on NMSE and cosine similarity, respectively, and compare our scheme to different CSI acquisition methods. In this experiment, the CSI acquisition period is set to $T_{ca}=2.5$~ms, implying that the pilot density for pilot-based channel estimation is $\alpha=0.66$ and the iteration number of our method is $K=7$. It is observed that our proposed method achieves the lowest NMSE and the highest cosine similarity among all the schemes. Another observation is that the performance of our proposed scheme and of the KNN-based channel inference scheme is almost invariant to SNR. This is because sensing-aided methods estimate CSI directly from environmental sensing data rather than relying on noisy pilot observations, which makes them robust to time-variant SNR levels. Although the KNN-based method achieves lower computational latency (as shown in Fig.~\ref{fig:Latency}), the experimental result seen in Fig. \ref{fig:Accuracy_over_SNR}(a) and Fig. \ref{fig:Accuracy_over_SNR}(b) demonstrates that its estimation quality is much worse than that of our scheme, since it only exploits user location and ignores the real-time dynamics of the wireless environment. By contrast, for all pilot-based channel estimation schemes, the performance improves upon increasing the SNR, as higher SNR leads to reduced pilot contamination by channel noise. However, in the high-SNR regime, their performance becomes limited by the number of available pilots, rather than by pilot quality. 
  
	Fig.~\ref{fig:Accuracy_over_Ratio}(a) and Fig.~\ref{fig:Accuracy_over_Ratio}(b) plot the NMSE and cosine similarity versus the CSI acquisition time. In this experiment, the SNR is fixed at 10~dB, and the CSI acquisition period $T_{ca}$ varies from 1.25~ms to 2.5~ms, corresponding to the pilot density $\alpha$ ranging from 0.11 to 0.66 for pilot-based channel estimation and the number of numerical integration steps $K$ increasing from 2 to 7 in our scheme. The results show that the proposed scheme consistently outperforms all benchmark methods across the considered range of CSI acquisition periods. For the KNN-based method, the inference time lies at the lower bound of the range considered (i.e., $T_{ca}=1.25$ ms), hence increasing the CSI acquisition period does not contribute to further performance gains. By contrast, all pilot-based channel estimation schemes benefit from a longer CSI acquisition period, since more pilots can be transmitted for reducing the channel estimation error.

    \begin{figure}[t]
    	\centering
    	\includegraphics[width = 0.93\linewidth]{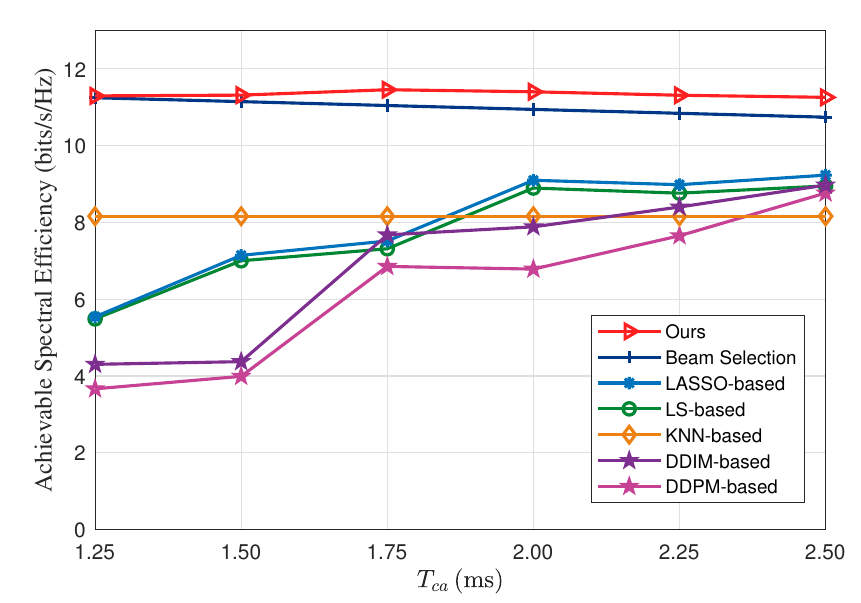}
    	\caption{The impact of CSI acquisition period $T_{ca}$ on achievable SE under the SNR = 10 dB. The numerical integration step $K$ varies with $T_{ca}$.}
    	\label{fig:Rate_over_Ratio}
    \end{figure}  
    
	\subsection{Evaluation on Achievable Spectral Efficiency}
	In Fig.~\ref{fig:Rate_over_SNR}, we evaluate the impact of SNR on the achievable SE for the hybrid beamforming task under different channel acquisition schemes, including both pilot-based channel estimation and sensing-aided channel inference methods. The channel acquisition period is set to $T_{ca}$ = 2.5 ms. As shown in the figure, the proposed method consistently outperforms all benchmark schemes across the SNR range considered. Compared to conventional pilot-based channel estimation schemes (i.e., LS, LASSO, DDPM and DDIM), the performance gain arises from two factors, namely improved channel estimation accuracy and the ability to reuse the channel acquisition time for data transmission. This indicates that the overall SE can be enhanced by reducing the pilot overhead through computation. Moreover, the SE improvement over the KNN-based channel inference scheme arises from more accurate channel inference in dynamic environments enabled by sensing environmental variations. While the multimodal sensing-aided beam selection method can also adapt to environmental dynamics, our scheme leverages the full CSI and thus yields more precise and stable beamforming decisions, leading to further SE gains.
   
    Fig.~\ref{fig:Rate_over_Ratio} illustrates the impact of the CSI acquisition period $T_{ca}$ on the achievable SE, where the SNR is fixed at 10~dB. The SE of the proposed scheme first increases and then shows a slight degradation as $T_{ca}$ becomes larger. The initial improvement comes from more accurate CSI enabled by increasing the number of numerical integration steps $K$. However, once the accuracy saturates at large $T_{ca}$ (see Fig.~\ref{fig:Accuracy_over_Ratio}), the loss caused by outdated CSI begins to dominate the performance, while the improved CSI accuracy only provides marginal SE gain. By contrast, the SE of the multimodal sensing-aided beam selection scheme decreases almost linearly with $T_{ca}$, since longer acquisition periods cause a larger portion of the data transmission to rely on outdated CSI. The achievable SE of the KNN-based scheme remains constant for all $T_{ca}$ values, as the beam selection does not change once the user’s location is fixed. This can also be confirmed from \eqref{eq:average_rate_ci}. For the pilot-based channel estimation schemes (i.e., LS, LASSO, DDPM and DDIM), within the smaller $T_{ca}$ regime, the SE increases with the improved channel estimation accuracy, but eventually saturates once the channel estimation is sufficiently accurate.

	\begin{figure}[t]
		\centering
		\includegraphics[width = 0.93\linewidth]{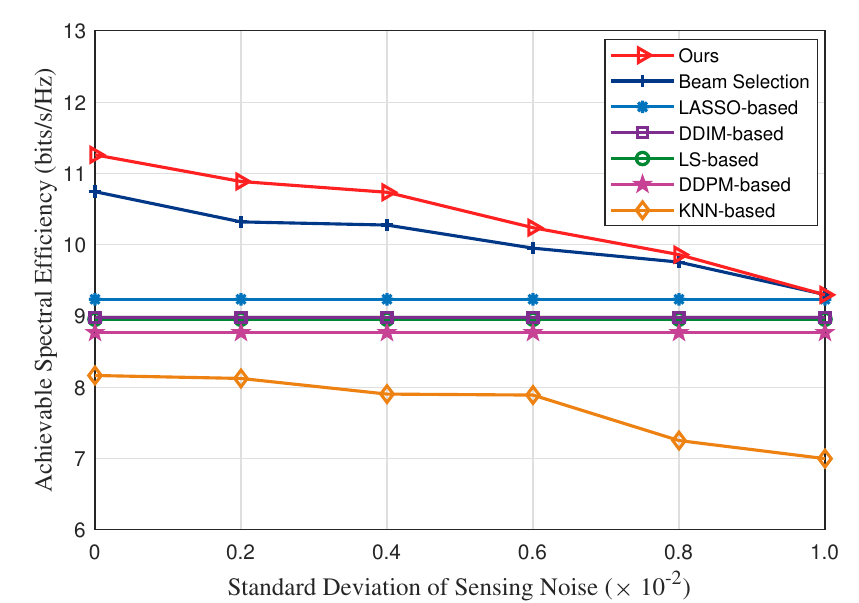}
		\caption{The impact of sensing degradation on achievable SE under the SNR = 10 dB and the CSI acquisition period $T_{ca}$ = 2.5 ms.}
		\label{fig:Rate_over_Error}
	\end{figure}  
	In Fig.~\ref{fig:Rate_over_Error}, we evaluate the impact of sensing degradation on the achievable SE, where the SNR is fixed at 10~dB and the CSI acquisition period is set to $T_{ca}=2.5$~ms. To provide a unified stress test that approximates the aggregate effect of possible sensing degradations, Gaussian noise with different standard deviations is added to each normalized sensing modality. As shown in the figure, the proposed method experiences a decrease in achievable SE as the sensing noise level increases, since noisy sensing inputs degrade the CSI estimation accuracy and thus lower the beamforming performance. However, it still consistently outperforms the benchmark schemes across the considered noise range, which confirms the robustness of the proposed sensing-assisted CSI inference framework. By contrast, the pilot-based schemes remain unchanged, as they rely on pilot measurements rather than sensing observations. The KNN-based scheme and the sensing-assisted beam selection scheme both degrade as the sensing noise increases, since location perturbations may cause mismatched nearest-neighbor channel path parameters, while noisy sensing modalities may disturb the beam-region decision. Moreover, the performance gap between the proposed method and the sensing-assisted beam selection scheme becomes smaller under larger sensing noise, since inferring high-dimensional and continuous CSI is more sensitive to input perturbations than predicting low-dimensional and discrete beam indices.

    \section{Conclusions}
    A novel framework was conceived for environment-aware channel inference that estimates the complete CSI directly from multimodal sensing data without relying on pilot signaling. By formulating the channel inference task as a cross-modal flow matching problem, the proposed method bridges the distributions of multimodal sensing inputs and channel representations through a learned continuous flow. A stochastic encoder and a neural velocity field were jointly optimized using the conditional flow matching and modality alignment losses to ensure tractable learning. The channel inference is represented as an ODE integration process, where a second-order numerical scheme enables real-time and model-free CSI estimation. System-level evaluations verify that the proposed framework consistently outperforms both pilot-based and existing sensing-based baselines in estimation accuracy and in downstream beamforming tasks.

	This work serves as an important step toward environment-aware communications. Several important directions remain for future research. First, improving cross-environment robustness is essential for practical deployment, which calls for both physics-informed neural modeling with explicit propagation priors and meta-learning-based fast adaptation to unseen scenes with limited additional data. Second, a further step toward real-world systems is to move beyond the ideal synchronized setting to asynchronous sensing-assisted channel inference, where state tracking and prediction mechanisms, together with lightweight pilot-based correction, will be important for compensating sensing staleness in high-mobility scenarios. Finally, distributed multimodal sensing offers a promising extension toward multi-view collaborative inference \cite{lyu2026quantizationawarecollaborativeinferencelarge}, which can improve the robustness of channel inference in more complex environments.
	
	\bibliography{Reference}
	
	\begin{IEEEbiography}
		[{\includegraphics[width=1in,height=1.25in,clip,keepaspectratio]{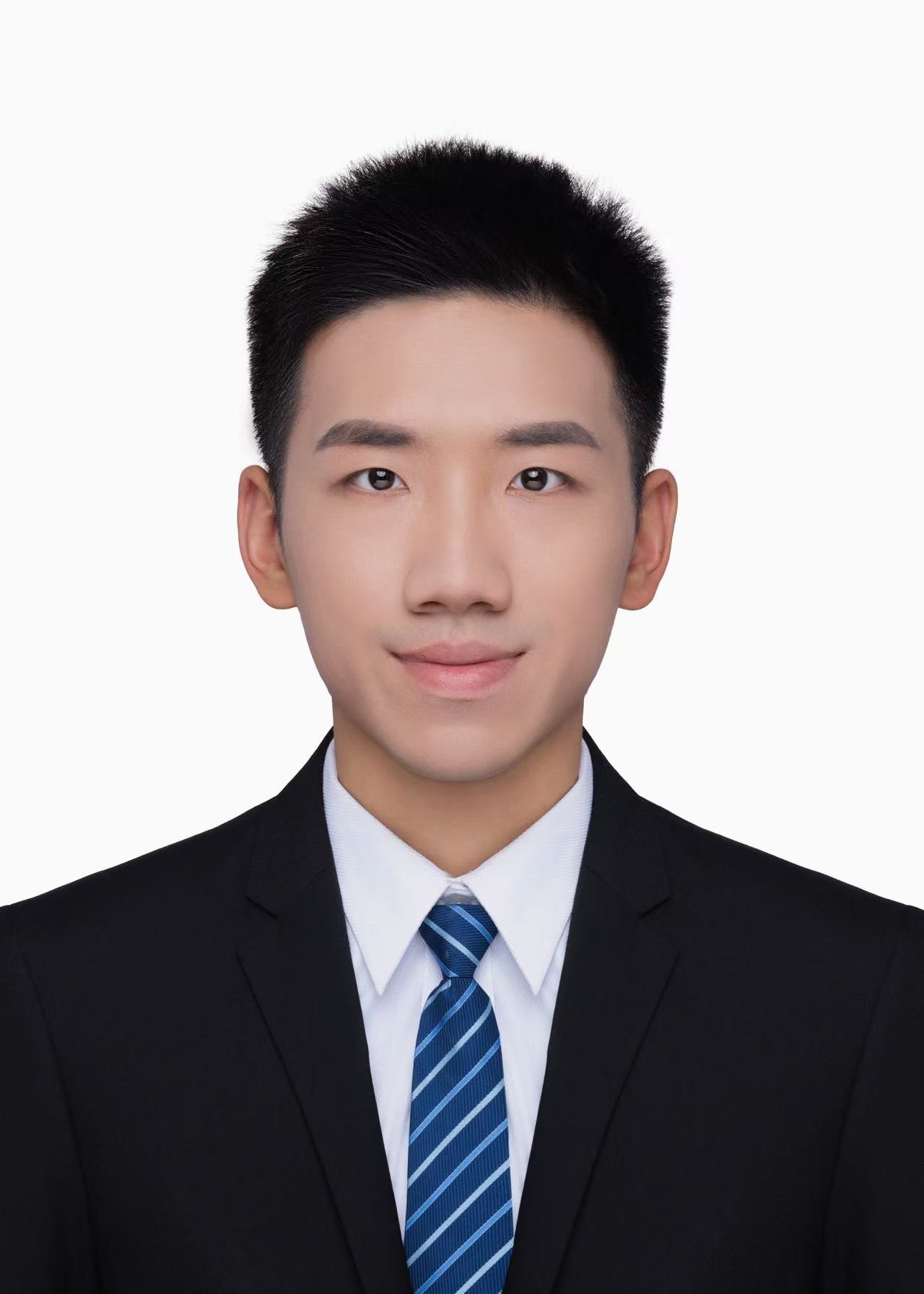}}] {Guangming Liang} (Graduate Student Member, IEEE) received the B.Eng. degree from Sun Yat-sen University (SYSU), Guangzhou, China, in 2021 and the M.Sc. degree from the University of Electronic Science and Technology of China (UESTC), Chengdu, China, in 2024. He is currently pursuing the Ph.D. degree with the School of Computing Science, University of Glasgow, Glasgow, U.K. His Ph.D. study is supported by the Graduate School Scholarship from the College of Science and Engineering, University of Glasgow, which covers full international tuition fees and a doctoral stipend. His current research interests include generative AI for wireless communications, multimodal sensing in wireless networks, and 6G digital twin networks. He received several honors and awards, including the Outstanding Graduate Awards from SYSU (2021), UESTC (2024), and Sichuan Province (2024), the Outstanding Master’s Thesis Award from UESTC (2024), as well as the UKRI Student Travel Grant (2026). He has served as a Technical Program Committee member for several prestigious IEEE conferences, including IEEE ICC, GLOBECOM and VTC.
	\end{IEEEbiography}

	\begin{IEEEbiography}
		[{\includegraphics[width=1in,height=1.25in,clip,keepaspectratio]{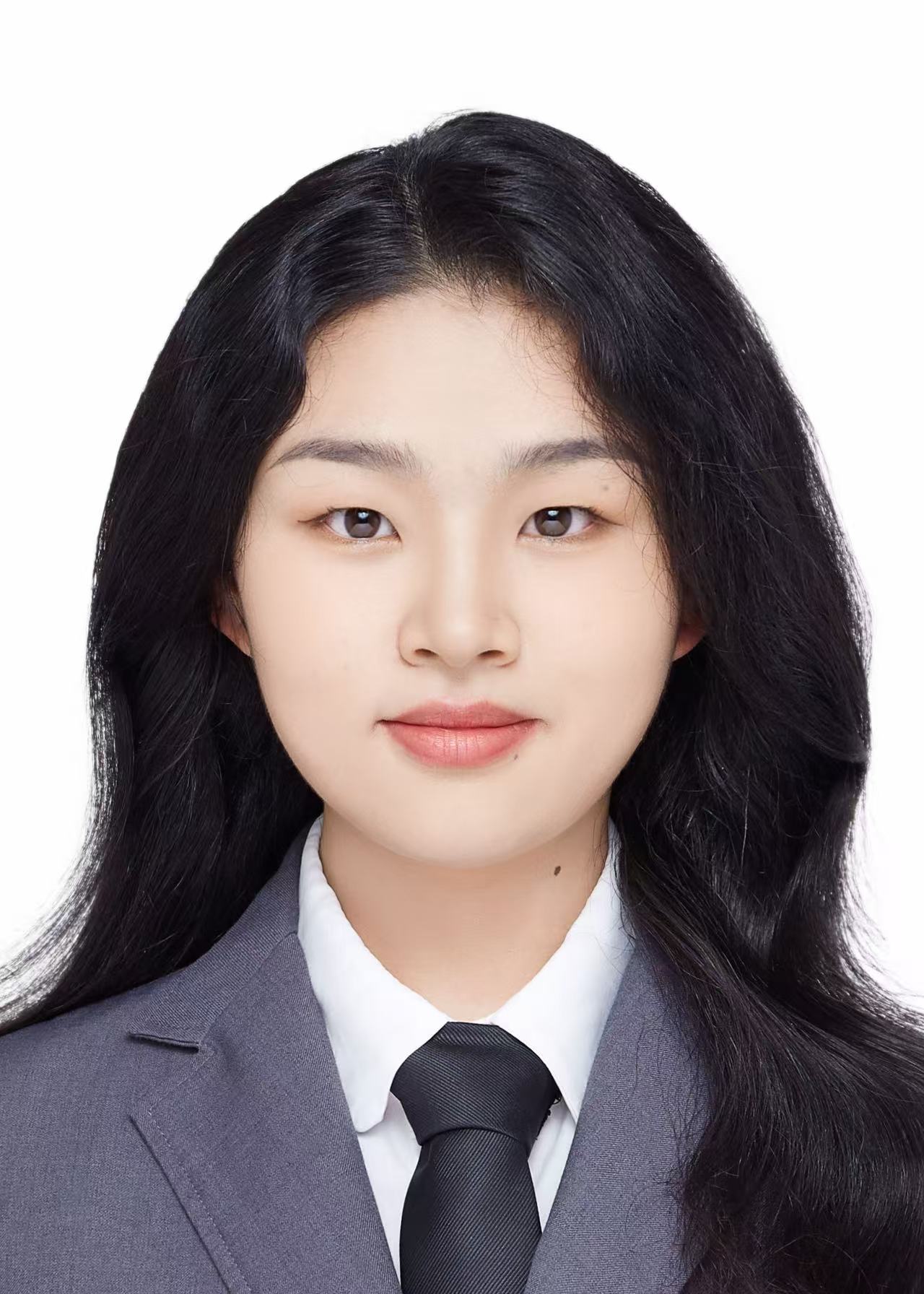}}] {Mingjie Yang}  (Graduate Student Member, IEEE) received her B.Eng. and M.Sc. degree from Xidian University in 2020 and 2023, respectively. She is currently pursuing the Ph.D. degree with the School of Computing Science, University of Glasgow, Glasgow, U.K. Her Ph.D. study is supported by a fully funded international doctoral scholarship from the College of Science and Engineering, University of Glasgow. Her current research interests include information-theoretic foundations of wireless sensing, communication-efficient sensing data transmission, and CSI-based sensing in cellular networks. She has served as a Technical Program Committee member for several prestigious IEEE conferences, including IEEE ICC and GLOBECOM.
	\end{IEEEbiography}	
    
    \begin{IEEEbiography}
    	[{\includegraphics[width=1in,height=1.25in,clip,keepaspectratio]{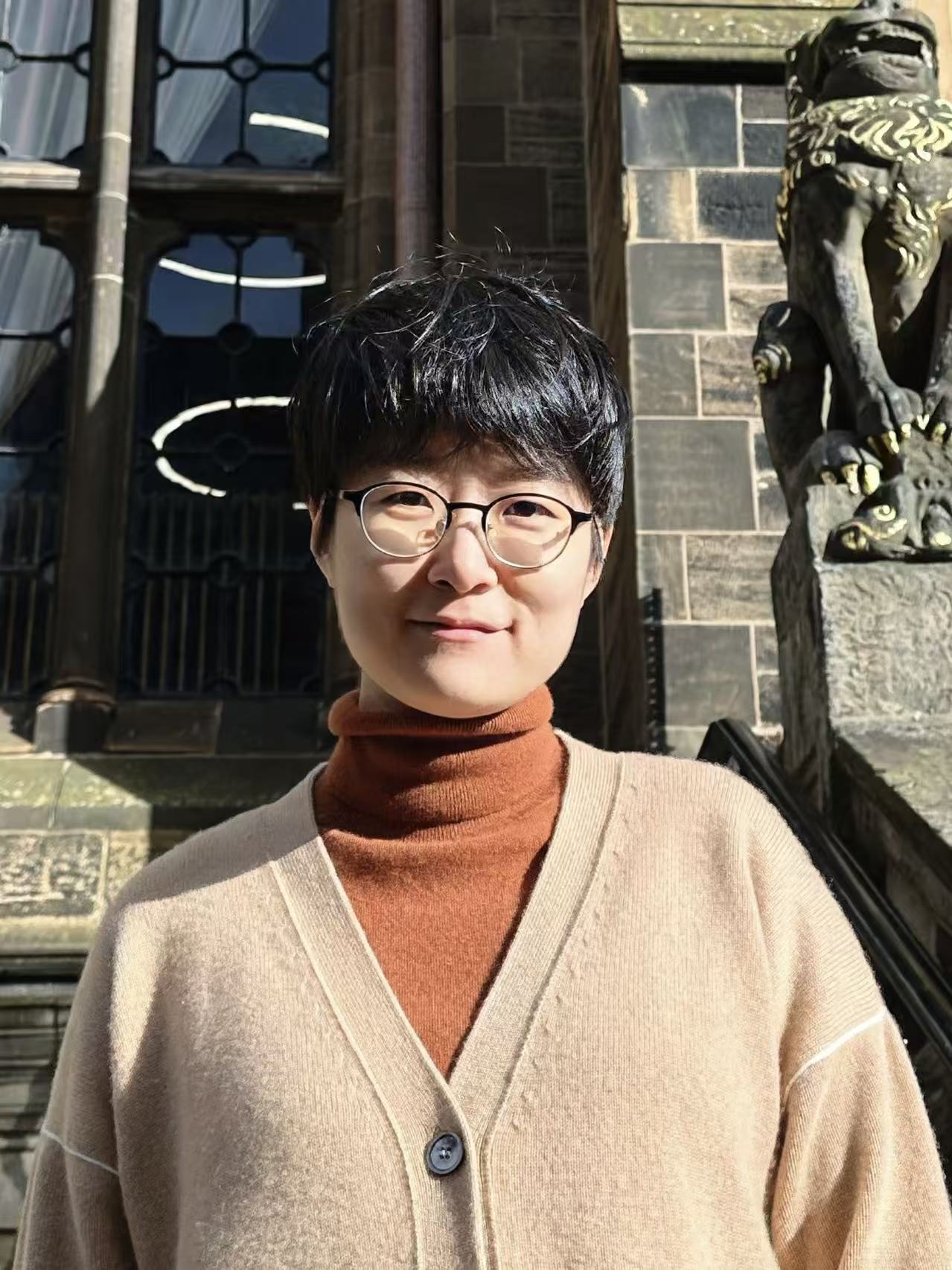}}] {Dongzhu Liu} (Member, IEEE) received the B.Eng. degree from the University of Electronic Science and Technology of China (UESTC) in 2015, and the Ph.D. degree from The University of Hong Kong in 2019. From 2019 to 2021, she was a Postdoctoral Research Associate with the Department of Engineering, King’s College London. She is currently a Lecturer (Assistant Professor) with the School of Computing Science, University of Glasgow. Her research interests include edge AI, wireless networks, multimodal sensing, and distributed Bayesian learning.  Dr. Liu has authored or coauthored a book chapter in Machine Learning and Wireless Communications and papers in IEEE Journal on Selected Areas in Communications, IEEE Transactions on Wireless Communications, and IEEE Communications Magazine. She is an Associate Editor for IEEE Transactions on Mobile Computing and IEEE Transactions on Machine Learning in Communications and Networking. She has served as a Co-Chair of workshops on task-oriented and generative communications for 6G at IEEE ICC, GLOBECOM, WCNC, and PIMRC, and as a Technical Program Committee member for IEEE ComSoc flagship conferences in wireless communications.
    \end{IEEEbiography}

	\begin{IEEEbiography}[{\includegraphics[width=1in,height=1.25in,clip,keepaspectratio]{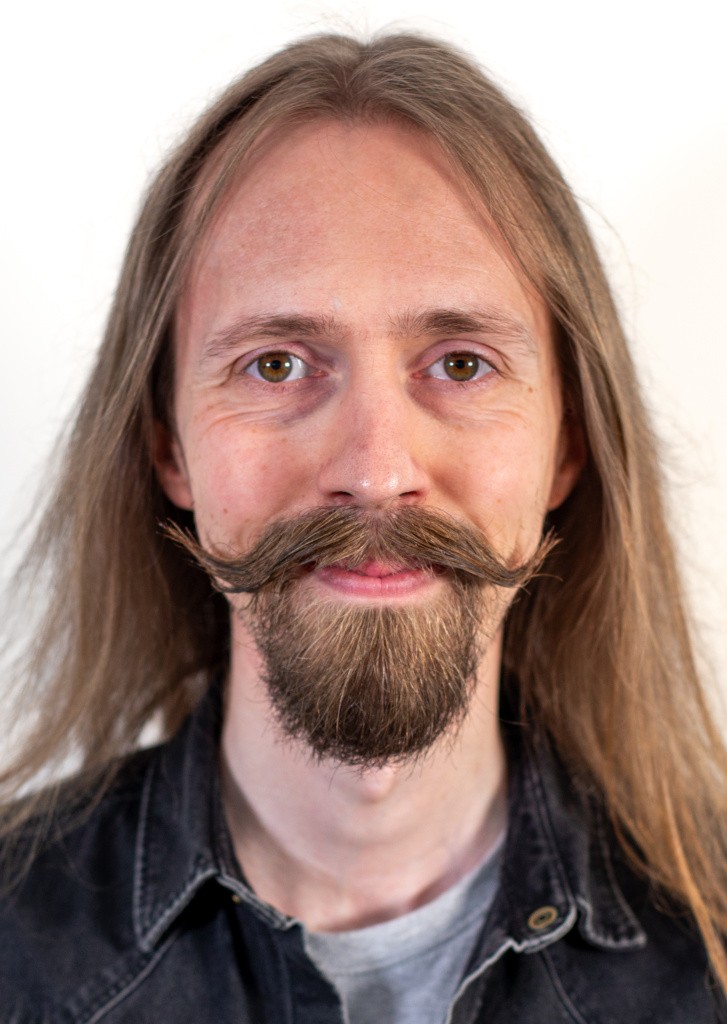}}]{Paul Henderson} received a BA in Mathematics from the University of Cambridge in 2009, MSc in Artificial Intelligence from the University of Edinburgh in 2010, and PhD in Informatics from the University of Edinburgh in 2018. He is a Senior Lecturer (Associate Professor) in Machine Learning at the University of Glasgow, where he specializes in generative AI for visual data, probabilistic machine learning, and 3D computer vision, including applications in the physical sciences and healthcare.
	\end{IEEEbiography}

	\begin{IEEEbiography}
		[{\includegraphics[width=1in,height=1.25in,clip,keepaspectratio]{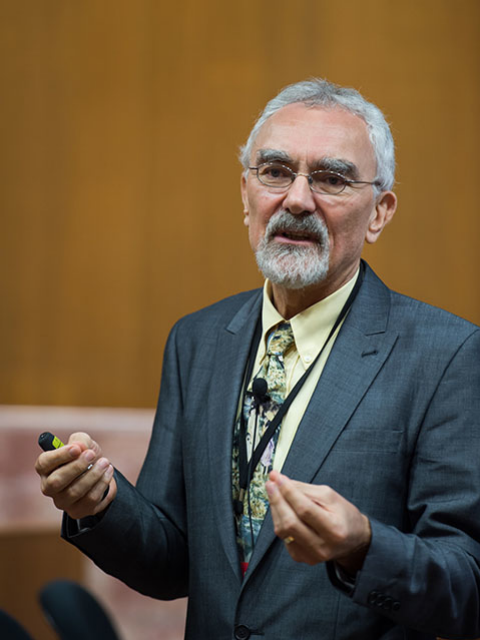}}] {Lajos Hanzo}  (Life Fellow, IEEE) received Honorary Doctorates from the Technical University of Budapest (2009) and Edinburgh University (2015). He is a Foreign Member of the Hungarian Academy of Sciences, Fellow of the Royal Academy of Engineering (FREng), of the IET, of EURASIP and holds the IEEE Eric Sumner Technical Field Award. He has published 2000+ contributions at IEEE Xplore, 20 Wiley-IEEE Press books and has helped the fast-track career of 126 PhD students. He holds the Chair of Telecommunications and directs the research of Next-Generation Wireless at the University of Southampton, UK.
	\end{IEEEbiography}

	\end{document}